\documentstyle[12pt]{article}
\begin{document}
{\hbox to\hsize{November 1995 \hfill IASSNS-HEP 95/99}}\par
{\hbox to\hsize{hep-ph/9512389 \hfill BA-95-55}}\par
\begin{center}
{\Large \bf {Realistic Quark and Lepton Masses \\[.1in]
Through SO(10) Symmetry}}\footnote{Work supported in part by
the Department of Energy}\\[.3in]
{\bf K.S. Babu}\footnote{E-mail: babu@sns.ias.edu}\\[.05in]
{\it School of Natural Sciences \\
Institute for Advanced Study\\
Princeton, NJ 08540}\\[.15in]
and \\[.1in]
{\bf S.M. Barr}\footnote{E-mail: smbarr@bartol.udel.edu}\\[.05in]
{\it Bartol Research Institute \\
University of Delaware\\
Newark, DE 19716}\\[.3in]
\end{center}
\begin{abstract}

In a recent paper a model of quark and lepton masses
was proposed. Without any family symmetries almost
all the qualitative and quantitative features of
the quark and lepton masses and Kobayashi-Maskawa
mixing angles are explained, primarily as consequences
of various aspects of $SO(10)$ symmetry.  Here the
model is discussed in much greater detail.  The threefold
mass hierarchy as well as the relations $m_\tau^0
\simeq m_b^0, ~m_\mu^0 \simeq 3m_s^0,~ m_e^0 \simeq {1 \over 3}
m_d^0, ~m_u^0/m_t^0 \ll m_d^0/m_b^0, ~{\rm tan}\theta_C \simeq
\sqrt{m_d^0/m_s^0},~ V_{cb} \ll \sqrt{m_s^0/m_b^0}$ and
$V_{ub} \sim V_{us}V_{cb}$ follow from a simple Yukawa
structure at the unification scale.  The model
also gives definite predictions for $\tan \beta$,
the neutrino mixing angles, and proton decay branching
ratios.  The $(\nu_\mu-\nu_\tau)$ mixing angle is typically
large, tan$\beta$ is close to either $m_t^0/m_b^0$ or
$m_c^0/m_s^0$, and proton decay is in the observable range,
but there is a group theoretical suppression factor in the rate.

\end{abstract}
\newpage

\baselineskip = .30in

\section{Introduction}
In a recent paper$^1$ we proposed a model of quark and lepton masses
based upon the gauge group $SO(10)$. In that paper it was shown how
most of the features of the fermion masses and mixings can be
understood to be consequences of various aspects of $SO(10)$
invariance rather than of some family symmetry or arbitrarily
imposed texture.

In this paper we go into greater detail. In Section 2 the
basic ideas of the model are explained and its structure is
reviewed in detail. The results of numerical fits are
presented in Section 3. Section 4 deals with the predictions
of the model for $\tan \beta$ and the neutrino mixing angles.
Predictions for proton-decay branching ratios are also briefly
discussed there (for details see Ref. (2)).
In Section 5 various
technical issues relating to the Higgs sector, symmetry breaking,
and the naturalness of the gauge hierarchy are examined.
Finally, in Section 6 certain alternative possibilities
for variant models are discussed.  Various technical details are
summarized in three Appendices.

\section{Review of the Model}

\noindent
{\large\bf (a) The Root Model}

The model we shall study is based on $SO(10)$, and though
supersymmetry is not essential for its account of quark
and lepton masses it shall be assumed because of the gauge
hierarchy. The ``matter" fields (quarks and leptons) are
contained in three spinors (${\bf 16}_i$, i = 1,2,3), which
are the ``families", and a real (in the group theory sense)
set of additional representations. These latter consist
of a pair of family and anti-family (${\bf 16} + \overline{{\bf 16}}$),
and a pair of vectors (${\bf 10} + {\bf 10}'$). In the ``long
version" of the model there is also a pair of adjoints (${\bf 45}
+ {\bf 45}'$). The Higgs multiplets which couple to the matter
are in a vector (${\bf 10}_H$), two adjoints (${\bf 45}_H +
\tilde{{\bf 45}}_H$), and a pair of spinors (${\bf 16}_H +
\overline{{\bf 16}}_H$). The short version of the model has
two sets of Yukawa terms in the superpotential, $W_{{\rm spinor}}$
and $W_{{\rm vector}}$. The long version has an additional set
of terms, $W_{{\rm adjoint}}$. These are given by
\begin{equation}
W_{{\rm spinor}} = M \overline{{\bf 16}} {\bf 16}
+ \sum_i b_i ({\bf 16}_i \overline{{\bf 16}}) {\bf 45}_H
+ \sum_i a_i ({\bf 16}_i {\bf 16}) {\bf 10}_H,
\end{equation}
\begin{equation}
W_{{\rm vector}} = d({\bf 10} \; {\bf 10}') \tilde{{\bf 45}}_H
+ \sum_i c_i ({\bf 16}_i {\bf 10}) {\bf 16}_H
+ \sum_i c_i^{\prime} ({\bf 16}_i {\bf 10}') {\bf 16}_H,
\end{equation}
\begin{equation}
W_{{\rm adjoint}} = f ({\bf 45} \; {\bf 45}') {\bf 45}_H
+ \sum_i e_i ({\bf 16}_i {\bf 45}) \overline{{\bf 16}}_H
+ \sum_i e_i^{\prime} ({\bf 16}_i {\bf 45}')
\overline{{\bf 16}}_H.
\end{equation}

Note that all three pieces of the Yukawa superpotential
have the same basic structure. In each there is one term
(the first) which couples together the pair of extra
representations and gives them superheavy mass, and two terms
(the second and third) which couple these extra representations
to the ordinary families. What is {\it not} allowed is a
direct mass term $\sum_i g_{ij} ({\bf 16}_i {\bf 16}_j) {\bf 10}_H$,
which would, unless there were fine tuning, tend to give
comparable masses to all the generations. All these three pieces,
in other words, have a kind of ``see-saw" structure in which the
ordinary families get masses through their mixing with the
extra fields.

The dominant contribution to the fermion mass matrices is assumed
to come from $W_{{\rm spinor}}$. (As will be seen in Section 5,
this is most simply explained as being a consequence of the
condition $\langle \overline{{\bf 5}}({\bf 10}_H) \rangle
\ll \langle \overline{{\bf 5}}({\bf 16}_H) \rangle$. Here and
throughout the notation ${\bf p}({\bf q})$ refers to a ${\bf p}$
of $SU(5)$ contained in a ${\bf q}$ of $SO(10)$.)
Diagrammatically, this contribution comes from the tree graph
shown in Fig. 1. Its approximate form can be read off from that
figure.
\begin{equation}
W_0 \cong \sum_{ij} a_i b_j \frac{\langle {\bf 10}_H \rangle
\langle {\bf 45}_H \rangle} {M} ({\bf 16}_i {\bf 16}_j).
\end{equation}

\noindent
Defining $\hat{a}_i = a_i/a$, $\hat{b}_i = b_i/b$ ($a= |\vec{a}|,
b=|\vec{b}|$); writing the VEV
of the ${\bf 45}_H$ as $\langle {\bf 45}_H \rangle =
\Omega Q$, where $\Omega$, like $M$, is of order $M_{{\rm GUT}}$,
and where $Q$ is a linear combination of $SO(10)$ generators;
and defining $T \equiv \frac{b \Omega}{M}$; one has
\begin{equation}
W_0 \cong a T \langle {\bf 10}_H \rangle \sum_{ij}
\left[ \hat{a}_i \hat{b}_j Q_{({\bf 16}_j)} \right]({\bf 16}_i
{\bf 16}_j).
\end{equation}

\noindent
Then for charge-$(\frac{2}{3})$ quarks the mass matrix, $U_{ij}$
is given by
\begin{equation}
U_{ij} \cong a T v \sum_{ij} \left[ \hat{a}_i \hat{b}_j Q_u
+ \hat{a}_j \hat{b}_i Q_{u^c} \right] u_i^c u_j,
\end{equation}

\noindent
with similar expressions for the mass matrices of the
charge-$(\frac{-1}{3})$ quarks ($D_{ij}$), charged leptons
($L_{ij}$), and the Dirac mass matrix of the neutrinos
($N_{ij}$). One can choose, without loss of generality,
the axes in family space so that $\hat{b}_i = (0,0,1)$
and $\hat{a}_i = (0, \sin \theta, \cos \theta)$. The
quantity $Q$ is a linear combination
of $SO(10)$ generators, with in general complex coefficients
since ${\bf 45}_H$ is a chiral superfield. There is a two-dimensional
space of such generators that commute with $SU(3)_c \times
SU(2)_L \times U(1)_Y$. We can thus choose to write
\begin{equation}
\begin{array}{c}
Q = 2 I_{3R} + \frac{6}{5} \epsilon \left( Y/2 \right),
\end{array}
\end{equation}

\noindent
where $I_{3R}$ is the third generator of $SU(2)_R$. Then
Eq. (6) and its analogues take the form

\begin{equation}
U_0 = aTv \left( \begin{array}{ccc}
0 & 0 & 0 \\
0 & 0 & Q_u \sin \theta/N_u \\
0 & Q_{u^c} \sin \theta /N_{u^c} & (Q_{u^c} + Q_u)
\cos \theta / N_{u^c} N_u
\end{array} \right),
\end{equation}

\begin{equation}
D_0 = aTv' \left( \begin{array}{ccc}
0 & 0 & 0 \\
0 & 0 & Q_d \sin \theta/N_d \\
0 & Q_{d^c} \sin \theta /N_{d^c} & (Q_{d^c} + Q_d)
\cos \theta / N_{d^c} N_d
\end{array} \right),
\end{equation}

\begin{equation}
L_0 = aTv' \left( \begin{array}{ccc}
0 & 0 & 0 \\
0 & 0 & Q_{l^-} \sin \theta/N_{l^-} \\
0 & Q_{l^+} \sin \theta /N_{l^+} & (Q_{l^+} + Q_{l^-})
\cos \theta / N_{l^+} N_{l^-}
\end{array} \right),
\end{equation}

\begin{equation}
N_0 = aTv \left( \begin{array}{ccc}
0 & 0 & 0 \\
0 & 0 & Q_{\nu} \sin \theta/N_{\nu} \\
0 & Q_{\nu^c} \sin \theta /N_{\nu^c} & (Q_{\nu^c} + Q_{\nu})
\cos \theta / N_{\nu^c} N_{\nu},
\end{array} \right),
\end{equation}

\noindent
where
\begin{equation}
\begin{array}{ccccl}
Q_u & = & Q_d & = & \frac{1}{5} \epsilon \\
& & Q_{u^c} & = & -1 - \frac{4}{5} \epsilon \\
& & Q_{d^c} & = & 1 + \frac{2}{5} \epsilon \\
Q_{l^-} & = & Q_{\nu} & = & -\frac{3}{5} \epsilon \\
& & Q_{l^+} & = & 1 + \frac{6}{5} \epsilon \\
& & Q_{\nu ^c} & = & -1
\end{array}
\end{equation}

\noindent
The factors $N_f \equiv \sqrt{1 + T^2 |Q_f|^2}$ come
from doing the algebra exactly rather than evaluating
the lowest order graph in Fig. 1. [The
linear combination of ${\bf 16}$'s which has a
Dirac mass with $\overline{{\bf 16}}$ and is
thus superheavy is clearly seen from Eq. (1) and
the definition of $T$ to be $({\bf 16} + T Q_{{\bf 16}_3}
{\bf 16}_3)/ \sqrt{1 + T^2 |Q_{{\bf 16}_3}|^2}$. So the
superheavy $u$, for example, is $(u_{{\bf 16}} +
T Q_u u_{{\bf 16}_3})/ \sqrt{1 + T^2 |Q_u|^2}$, etc.
Finding the orthogonal (light) linear combinations and
writing the term $a_i {\bf 16}_i {\bf 16} {\bf 10}_H$
which appears in Eq. (1) in terms of them, one obtains the
exact expressions in Eqs. (8) -- (11).]

The striking feature of Eqs. (8) -- (11) is that the mass
matrices are rank 2. This is a consequence of `factorization';$^{3,4}$
that is, that the mass matrices do not come from a Yukawa coupling
coefficient that is a $matrix$ in family space but from a product of
Yukawa coefficients that are $vectors$ in family space. The rank
2 comes directly from the fact that $\sum_i (a_i {\bf 16}_i) {\bf 16} \;
{\bf 10}_H$ involves just two distinct linear combinations of the light
generations, namely $\hat{a}_i {\bf 16}_i$ and ${\bf 16}$. In
other words, the fact that there are two heavy generations is a
direct consequence of the fact that Yukawa terms are {\it bilinear}
in matter fields. (It is amusing to note that in this context
if there were less than three generations then there would be no
light family such as makes up the ordinary matter of our world.
Perhaps this is a partial answer to the famous question of Rabi.
Without the $\mu$ and the $\tau$ one would not have the electron!)

A second striking feature is that these equations provide
an explanation of the puzzling fact that the minimal $SU(5)$
prediction$^5$ $m_b^0 \cong m_{\tau}^0$ works so well while
the corresponding predictions for the lighter generations fail
badly. (Here and throughout, the superscript $^0$ refers to
a quantity evaluated at the GUT scale.) $m_b^0 \cong m_{\tau}^0$
is a consequence of the fact that $D_{33} \cong L_{33}$, which
follows from the
relation $Q_{d^c} + Q_d = Q_{l^+} + Q_{l^-}$. This relation, in
turn, is implied by the fact that both $d$ and $l$ get mass
from the same Higgs field, $H'$, so that $Q_{d^c} + Q_d + Q_{H'}
= Q_{l^+} + Q_{l^-} + Q_{H'} = 0$. On the other hand, $m_{\mu}^0
\cong - \frac{L_{32} L_{23}}{L_{33}}$ and $m_s^0 \cong
- \frac{D_{32} D_{23}}{D_{33}}$, so that $m_{\mu}^0/m_s^0
\cong \frac{L_{32} L_{23}}{D_{32} D_{23}} = \frac{Q_{l^+} Q_{l^-}}
{Q_{d^c} Q_d}$, which is a group-theoretical factor of order unity
but not equal to unity in general.

Another way to understand the fact that $m_b^0 \cong m_{\tau}^0$
while $m_s^0 \neq m_{\mu}^0$ is from the group-theoretical
structure of the term ${\bf 16}_i {\bf 16}_j {\bf 10}_H {\bf 45}_H$.
({\it Cf}. Eq. (4).) The ${\bf 10}_H \times {\bf 45}_H$ can be
contracted into a ${\bf 10}$, ${\bf 120}$, or ${\bf 320}$ of $SO(10)$.
${\bf 320}$ is not contained in ${\bf 16} \times {\bf 16}$ and
so does not couple. If $i = j = 3$ then the contraction ${\bf 120}$,
which must couple antisymmetrically in flavor, is also not allowed.
Hence only the contraction into a ${\bf 10}$ couples to ${\bf 16}_3
{\bf 16}_3$, and therefore the minimal $SU(5)$ relation
$D_{33} = L_{33}$ holds. But for $(i,j) = (2,3)$ or $(3,2)$ the
antisymmetric contraction into a ${\bf 120}$ also couples, and
thus the minimal $SU(5)$ relation does not hold for the second
generation.

Three more facts are explained if the assumption is made that
$\left| \epsilon \right| \ll 1$, {\it ie}. that $\langle
{\bf 45}_H \rangle$ points approximately in the $I_{3R}$
direction. First, since $\left| \frac{m_{\mu}^0}{m_s^0} \right|
\cong \left| \frac{Q_{l^+}Q_{l^-}}{Q_{d^c}Q_d} \right| =
3 \left| \frac{1+ \frac{6}{5} \epsilon}{1 + \frac{2}{5} \epsilon}
\right|$, the Georgi-Jarlskog relation$^6$ $m_{\mu}^0 \cong 3 m_s^0$
is explained. Second, since $Q_u$, $Q_d$, and $Q_{l^-}$ are all
proportional to $\epsilon$ ($u_L$, $d_L$, and $l^-_L$ are all
singlets under $SU(2)_R$),
the matrices $U$, $D$, and $L$ all become rank 1 in the limit
$\epsilon \rightarrow 0$. Thus the hierarchy between the second
and third generation masses is explained. It is remarkable that
a relation among generations is related to a direction in the
gauge group space. And, third, the angle $V_{cb}$ is of order
$\epsilon$. ($V_{cb}^0 \cong \left(\frac{U_{32}}{U_{33}} \right)
- \left( \frac{D_{32}}{D_{33}} \right) = \tan \theta \left[
\frac{Q_{u^c}}{Q_{u^c} + Q_u} - \frac{Q_{d^c}}{Q_{d^c} + Q_d} \right]
\cong  \tan \theta (Q_d + Q_u) = \frac{2}{5} \epsilon \tan \theta$.)
This explains the smallness of $V_{cb}$ compared to $\sqrt{m_s^0/
m_b^0}$, which is a puzzle in Fritzschian models.$^7$ In fact, since
$m_s^0/m_b^0 \cong \frac{1}{5} \epsilon \sin^2 \theta$, one has
for $\theta \sim 1$ that $V_{cb}^0 \sim 2 \left( \frac{m_s^0}
{m_b^0} \right)$, which is a true relation.

A further consequence of the structure of Eq. (4) is that the
Higgsino-mediated proton-decay amplitude
is proportional to$^2$ $\epsilon$. Thus the smallness of $\epsilon$
helps$^8$ solve the problem of excessively rapid proton decay
which tends to afflict supersymmetric grand unified models,$^9$
especially those with large $\tan \beta$. (It should be
noted that many SUSY GUT models based on $SO(10)$ give
$\tan \beta \cong m_t^0/m_b^0$.)  The branching ratio $Br[(p
\rightarrow K^+ \overline{\nu}_\mu)/(p \rightarrow \pi^+
\overline{\nu}_\mu)]$ is a calculable quantity in the model$^2$, but
it differs from that of minimal SUSY $SU(5)$.  Similalry, the charged
lepton mode $p \rightarrow K^0\mu^+$ which could become significant in
the large tan$\beta$ scenario has a different branching ratio than in
$SU(5)$. These distinctions could serve as testing grounds for the
high scale flavor structure.

In summary up to this point, the simple structure of $W_{{\rm spinor}}$
together with the assumption that $Q \sim I_{3R}$ has explained
or contributed to explaining six facts. And the single group-theoretical
assumption about the direction of $Q$ has played a role in four of these
explanations, a striking economy.
It should be noted that the direction $I_{3R}$ is not an arbitrary
one but a point of higher symmetry, so that certain simple
superpotentials can give an
adjoint Higgs a VEV in that direction. For example, a superpotential
with the terms $W = S^3 + S^2 + S A^2 + A^2$, where $A$ is a ${\bf 45}$
and $S$ is a ${\bf 54}$ has solutions for $A$ in the $I_{3R}$ and $B-L$
directions.$^{10}$
And a superpotential with terms $W= A^2 + (A^2)^2 + A^4$
has solutions in the $I_{3R}$, $B-L$
and $X$ directions$^{11}$.
($X$ is the $SU(5)$-singlet generator of $SO(10)$.)
It is also interesting that the Dimopoulos-Wilczek mechanism,$^{12}$ which
appears to be the only viable way to make the gauge hierarchy
natural in $SO(10)$, requires there to exist an adjoint VEV in the
$B-L$ direction.

At this point, more needs to be said about the factors $N_f \equiv
\sqrt{1 + T^2 |Q_f|^2}$. Note that for $T$ small, which corresponds
to small mixing between the ${\bf 16}_i$ and the ${\bf 16}$, these
factors become very close to unity and can be ignored. [$T$ cannot
be {\it too} small, however, since $m_t^0 \cong aTv \cos \theta$, and
therefore $a \stackrel{_>}{_\sim} 1/T$. For $a \gg 1$ perturbation
theory would break down above the GUT scale.] Even with $T$ of order
unity the $N_f$ become simple in the limit $\left| \epsilon \right|
\ll 1$, where $N_{u,d,l^-} = 1 + O(\epsilon^2)$, and $N_{u^c, d^c,
l^+} = \sqrt{1 + T^2} + O(\epsilon)$. Effectively, then, there is
for small $\epsilon$ only a single parameter, $N \equiv \sqrt{1 + T^2}
\sim 1$, introduced by these mixing factors.
This parameter, $N$, plays no significant role. For without it there
would be three parameters ($\epsilon$, $\theta$, and $\tan \beta$) to
fit the five mass ratios and one mixing angle of the second and third
generations. That means there would be three quantitative ``predictions",
which are (for the small $\epsilon$ case) $m_{\tau}^0 \cong m_b^0$,
$m_{\mu}^0 \cong 3 m_s^0$, and $m_c^0/m_t^0 \cong m_s^0/m_b^0$
(about which more later). But one easily sees that the factors
of $1/N$ appearing in the (32) and (33) elements of all the
mass matrices approximately cancel out in all these relations and
thus leave them unaffected. As for the two qualitative ``predictions",
namely that $V_{cb}$ and the ratios of second to third generation
masses are of order $\epsilon$ and hence small, these are clearly
unaffected by the presence of $N$, which is of order unity.

The simple structure of $W_{{\rm spinor}}$ (Eq. (1)),
as we have just seen,
explains in a highly economical fashion most of the features of the masses
and mixings of the two heavy generations, and indeed the very fact that
there are two heavy generations. Two facts about the heavy generations are
not explained. The first is that $m_t^0 \gg m_b^0$, which is just
equivalent in this model to the statement that $\tan \beta \gg 1$.
An explanation of this must rely on an understanding of the Higgs sector.
If we regard this model as a model of the Yukawa sector, this question
is beyond its scope.
The second unexplained fact is the failure of the ``proportionality"
relation of $SO(10)$, $m_c^0/m_t^0 = m_s^0/m_b^0$, which is not
significantly broken by any effects discussed so far. Different
mechanisms for breaking this bad relation lead to different versions
of the model, see subsections (c) and (d) below.

\vspace{0.5cm}

\noindent
{\large\bf (b) The first generation: $W_{{\rm vector}}$}

The masses and mixings of the first generation arise from the next
layer of the model, given by $W_{{\rm vector}}$. As before, in the
case where the mixing of the ${\bf 16}_i$ with the ${\bf 10} +
{\bf 10}'$ is small the contribution of $W_{{\rm vector}}$ to
the light quark and lepton masses can be read off from a simple
graph, shown in Fig. 2. At first glance it might seem that many new
parameters are introduced by adding $W_{{\rm vector}}$. (In fact,
nine: the six Yukawa couplings, $c_i$ and $c_i'$, and three VEVs.)
But the important point is that in the limit of small mixing Fig. 2
gives a {\it flavor-antisymmetric} contribution to $D_{ij}$ and
$L_{ij}$. (There is no contribution to $U_{ij}$ as ${\bf 10}$ and
${\bf 10}'$ do not contain charge-($\frac{2}{3})$ quarks.)
The flavor-antisymmetry is obvious from Fig. 2, for under the
interchange ${\bf 10} \leftrightarrow {\bf 10}'$ the Yukawa
coupling ${\bf 10} \; \tilde{{\bf 45}}_H {\bf 10}'$ changes sign
because the adjoint of $SO(10)$ is an antisymmetric tensor,
while the flavor indices $i$ and $j$ are interchanged.
Moreover, if it is assumed that the $\tilde{{\bf 45}}_H$
acquires an $SU(5)$-invariant VEV, the contributions to $D_{ij}$
and $L_{ij}$ will be equal (up to a sign since $\delta D = \delta L^T$).
Thus {\it in the small mixing limit} (we will return to the more
general case later) there are really only {\it three} new
parameters introduced which we will call $c_{ij}$, ($i \neq j$),
where
\begin{equation}
c_{ij} = (c_i c_j' -  c_j c_i') \frac{M \langle {\bf 1}({\bf 16}_H)
\rangle}{abd \langle {\bf 45}_H \rangle \langle \tilde{{\bf 45}}_H
\rangle} \frac{ \langle \overline{{\bf 5}}({\bf 16}_H) \rangle}
{\langle \overline{{\bf 5}} ( {\bf 10}_H ) \rangle }.
\end{equation}

\noindent
It should be noted that if the $c_{ij}$ are all comparable
then $c_{23}$ is small compared to the other contributions to
the terms in which it appears ($D_{23}$, $D_{32}$, $L_{23}$, and
$L_{32}$) and can therefore be neglected.

The mass matrices then take the form (putting in also the values
of the $Q_f$ from Eq. (12))

\begin{equation}
U = aTv \left( \begin{array}{ccc}
0 & 0 & 0 \\
0 & 0 & \frac{1}{5} \epsilon \sin \theta/N_u \\
0 & -(1+ \frac{4}{5} \epsilon) \sin \theta /N_{u^c} & -(1+ \frac{3}{5}
\epsilon)
\cos \theta / N_{u^c} N_u
\end{array} \right),
\end{equation}

\begin{equation}
D = aTv' \left( \begin{array}{ccc}
0 & -c_{12} & -c_{13}/N_d \\
c_{12} & 0 & (\frac{1}{5} \epsilon \sin \theta -c_{23})/N_d \\
c_{13}/N_{d^c} & ((1+ \frac{2}{5} \epsilon) \sin \theta
+ c_{23})/N_{d^c} & (1+ \frac{3}{5} \epsilon)
\cos \theta / N_{d^c} N_d
\end{array} \right),
\end{equation}

\begin{equation}
L = aTv' \left( \begin{array}{ccc}
0 & c_{12} & c_{13}/N_{l^-} \\
-c_{12} & 0 & (-\frac{3}{5} \epsilon \sin \theta
+ c_{23})/N_{l^-} \\
-c_{13}/N_{l^+} & ((1 + \frac{6}{5} \epsilon) \sin \theta
-c_{23})/N_{l^+} & (1+ \frac{3}{5} \epsilon)
\cos \theta / N_{l^+} N_{l^-}
\end{array} \right),
\end{equation}

\begin{equation}
N = aTv \left( \begin{array}{ccc}
0 & 0 & 0 \\
0 & 0 & - \frac{3}{5} \epsilon \sin \theta/N_{\nu} \\
0 & - \sin \theta /N_{\nu^c} & -(1 + \frac{3}{5} \epsilon)
\cos \theta / N_{\nu^c} N_{\nu},
\end{array} \right),
\end{equation}

Several more features of the quark and lepton mass spectrum
are explained by this form. First, that $U_{ij}$ is still rank
2 corresponds to the fact that $m_u^0/m_t^0$ ($\approx 10^{-5}$)
is much smaller than $m_d^0/m_b^0$ ($\approx 10^{-3}$) and
$m_e^0/m_{\tau}^0$ ($\cong 0.3 \times 10^{-3}$).
Second, the antisymmetry of the contributions to the first
row and column of $D_{ij}$ implies that one has effectively the
form $\left( \begin{array}{cc} 0 & A \\ -A & B \end{array} \right)$
for the (12) block of that matrix (after diagonalizing the (23) block).
This is precisely the well-known form$^{13}$ that leads to the famous
relation$^{13,14}$
$\tan \theta_c = \sqrt{m_d^0/m_s^0}$. (There is no $\sqrt{m_u^0/m_c^0}$
contribution. Some other effect will generate a non-zero and very
small $m_u$, but there is no reason for this other effect to have a
form that makes $U_{11} \ll U_{12}, U_{21}$. Rather, if one supposes
that $U_{11}$, $U_{12}$, and $U_{21}$ are all of the same order then
the contribution from diagonalizing $U$ to $\tan \theta_c$ will be
$O(m_u^0/m_c^0) \sim 0.005$ instead of the usual $\sqrt{m_u^0/m_c^0}
\cong 0.07$. Thus this ``prediction" works better here than in the usual
scenarios.)

A third consequence of the forms given in Eqs. (15) and (16) is that
\begin{equation}
\det D = \det L.
\end{equation}

\noindent
Note that, remarkably, this is true for {\it any} values of
$\theta$, $\epsilon$,
and $c_{ij}$, if $T$ is small enough that the factors $N_f$ can
be taken to be one. And for any $T$, this equality holds for small $\epsilon$.
{}From this follows the well-known Georgi-Jarlskog relation$^6$
$m_e^0/m_d^0 \cong (m_{\mu}^0/m_s^0)^{-1} \cong \frac{1}{3}$.

Finally, if all the $c_{ij}$ are comparable then $V_{ub} \sim
V_{us} V_{cb}$, which is the true order of magnitude statement
which underlies the Wolfenstein parameterization.$^{15}$ (For
$V_{ub} \sim D_{31}/D_{33} \sim aTv'(c_{13}/m_b^0)$, $V_{us} \sim
aTv'(c_{12}/m_s^0)$, and $V_{cb} \sim m_s^0/m_b^0$, the last being
both empirically true and a consequence in the model of $\theta
\sim 1$ as noted earlier.)

\vspace{0.5cm}

\noindent
{\large\bf (c) The proportionality relation: The short model}

The structure described so far, given in Eqs. (1) and (2), gives
a satisfactory account of all the features of the quark and lepton
spectrum (the threefold hierarchy, the mixing angles, and the
mass ratios) with one glaring exception. The ratio $m_c^0/m_t^0$
is seemingly predicted to be equal to $m_s^0/m_b^0$, whereas
empirically it is about one-fifth of that. The most obvious explanation
of this anomaly would be that another additive contribution to
$U_{ij}$ exists which happens to approximately cancel $U_{23}$.
It is plausible that this might happen without notably changing
any of the other successful features of the model since $U_{23}$
is a small element and also does not affect the mixing angles.
This is an attrctive possibility. But it suffers from the apparent
difficulty that the new contribution to $U_{ij}$ must leave it
very nearly rank 2 in order not to produce a value of $m_u$ that
is too large. We have found a way to do this, which is discussed in
Section 6(b). This idea, however has certain drawbacks, discussed in
Section 6(b), that make it less attractive than the ideas we shall
now present. However, we cannot discount the possibility that a more
elegant solution along these lines may exist.

A remarkable fact about the model described so far, containing only
the Yukawa couplings in Eqs. (1) and (2), is that it already contains,
{\it without} any
further additions, a {\it multiplicative} correction
to the mass matrices that can break the bad proportionality
relation while leaving the other good relations largely intact.

We have emphasized that the forms of the mass matrices given in
Eqs. (14) --(17) are valid for small mixing. The exact expressions for the
$D$, $L$, and $N$ matrices including all mixing effects are

\begin{equation}
D = aTv' (I + \Delta_{d^c})^{-\frac{1}{2}} \left( \begin{array}{ccc}
0 & -c_{12} & -c_{13}/N_d \\
c_{12} & 0 & (\frac{1}{5} \epsilon \sin \theta -c_{23})/N_d \\
c_{13}/N_{d^c} & ((1+ \frac{2}{5} \epsilon) \sin \theta
+ c_{23})/N_{d^c} & (1+ \frac{3}{5} \epsilon)
\cos \theta / N_{d^c} N_d
\end{array} \right),
\end{equation}

\begin{equation}
L = aTv' \left( \begin{array}{ccc}
0 & c_{12} & c_{13}/N_{l^-} \\
-c_{12} & 0 & (-\frac{3}{5} \epsilon \sin \theta
+ c_{23})/N_{l^-} \\
-c_{13}/N_{l^+} & ((1 + \frac{6}{5} \epsilon) \sin \theta
-c_{23})/N_{l^+} & (1+ \frac{3}{5} \epsilon)
\cos \theta / N_{l^+} N_{l^-}
\end{array} \right)  (I + \Delta_{l^-}^T)^{-\frac{1}{2}},
\end{equation}

\begin{equation}
N = aTv \left( \begin{array}{ccc}
0 & 0 & 0 \\
0 & 0 & - \frac{3}{5} \epsilon \sin \theta/N_{\nu} \\
0 & - \sin \theta /N_{\nu^c} & -(1 + \frac{3}{5} \epsilon)
\cos \theta / N_{\nu^c} N_{\nu},
\end{array} \right) (I + \Delta_{\nu}^T)^{-\frac{1}{2}},
\end{equation}

\noindent
where
\begin{equation}
(\Delta_{d^c})_{ij} = C_i^* C_j + C_i^{* \prime} C_j^{\prime},
\end{equation}

\noindent
and where
\begin{equation}
(C_1,C_2,C_3) \equiv \left| \frac{\langle {\bf 1}({\bf 16}_H) \rangle}
{d \langle \tilde{{\bf 45}}_H \rangle } \right| (c_1,c_2, c_3/N_{d^c}),
\end{equation}

\noindent
with a similar expression for $C_i^{\prime}$ in terms of the $c'_i$.
The expression for $\Delta_{l^-}$ is of the same form with
$N_{l^-}$ instead of $N_{d^c}$, and $\Delta_{\nu} = \Delta_{l^-}$
by $SU(2)_L$.  See Appendix A for the derivation.
The expression for $U_{ij}$ given in Eq. (14) is exact as it stands.

The matrices $\Delta_f$ characterize the mixing between the ${\bf 16}_i$
and the ${\bf 10} + {\bf 10}'$, just as $T Q_{{\bf 16}_i}$ characterizes
the amount of mixing between the ${\bf 16}_i$ and the ${\bf 16}$. The
factors $(I + \Delta_f)^{-\frac{1}{2}}$ are completely analogous
to the factors of $(1 + T^2 |Q_f|^2)^{-\frac{1}{2}} \equiv N_f^{-1}$.
For small mixing these factors are close to the identity and can be
neglected. Even if the $\Delta_f$ are not small, most of the
quantitative and qualitative successes of the model are only
slightly affected. Obviously the fact that $m_u \approx 0$ and the
threefold hierarchy are among these. Moreover, if $N_f \cong 1$ then
for arbitrarily large $\Delta_f$ the relation $\det D \cong \det L$
continues to hold, since then $\Delta_{d^c} \cong \Delta_{l^-}^T$.

Consider a case in which

\begin{equation}
(I + \Delta_{d^c})^{-\frac{1}{2}} \cong (I + \Delta_{l^-})^{-\frac{1}{2}}
\cong \left(
\begin{array}{ccc}
1 & & \\
& 1 & \\
& & \delta
\end{array}
\right),
\end{equation}

\noindent
where $\delta < 1$. From Eqs. (19) and (20) it can be seen that the main
effect of this factor is to suppress the $\tau$ and $b$ masses by a factor
of approximately $\delta$. This suppression can be understood
simply and intuitively in the following way. $\delta \ll 1$ arises,
as can be seen from Eqs. (22) and (23), from a large value of $c_3$ or $c_3'$
compared to $d \langle \tilde{{\bf 45}_H} \rangle/ \langle {\bf 1}({\bf 16}_H)
\rangle$. From Eq. (2) one sees that there is a superheavy mass term
${\bf 5}({\bf 10})[(d\langle \tilde{{\bf 45}}_H \rangle ) \cdot
\overline{{\bf 5}} ( {\bf 10}' ) + ( c_3 \langle {\bf 1} ( {\bf 16}_H )
\rangle ) \cdot \overline{{\bf 5}} ( {\bf 16}_3) ]$. Large $c_3$,
therefore, corresponds to the superheavy linear combination being
approximately $\overline{{\bf 5}}({\bf 16}_3)$ and the orthogonal, light
linear combination that is the third generation being approximately
$\overline{{\bf 5}} ({\bf 10}')$. That means that the term
$a \cos \theta \; \overline{{\bf 5}}( {\bf 16}_3) {\bf 10} ( {\bf 16} )
\langle \overline{{\bf 5}} ( {\bf 16}_H ) \rangle$ in Eq. (1) gives
a contribution mostly to the {\it superheavy} $\overline{{\bf 5}}$
rather
than to the third generation $\overline{{\bf 5}}$ which contains
$b^c$ and $\tau^-$. Thus the masses of $b$ and $\tau$ are suppressed
by a factor of approximately $\delta$.

It is easily seen by substituting Eq. (24) into Eqs. (19) and (20)
that the masses of $b$ and $\tau$ are multiplied approximately
by $\delta$ while the masses of $s$ and $\mu$ are not much
affected. Thus the ratio $m_s^0/m_b^0$ is enhanced relative to
$m_c^0/m_t^0$ by a factor of $\delta^{-1}$. Hence $\delta$ should be
about $\frac{1}{5}$. At the same time, the near cancellation between the
(23) mixing angles in the up and down sectors that is responsible
for the smallness of $V_{cb}$ is not disturbed, since the ratio
$D_{32}/D_{33}$ is left unchanged.

What this shows is that the factors $(I + \Delta)^{-\frac{1}{2}}$
in Eqs. (19) and (20) allow a fit to be made to $(m_c^0/m_t^0)/
(m_s^0/m_b^0)$ without greatly disturbing the qualitative and quantitative
successes of the model. However, there are certain prices to be paid
for this method of breaking the proportionality relation. First,
$\theta$ must be somewhat small. Since $V_{cb} \cong \frac{2}{5}
\epsilon \tan \theta$ and $m_s^0/m_b^0 \cong \frac{1}{5} \epsilon
\sin^2 \theta / \delta$, it follows that $\theta \cong 2 \delta
\frac{m_s^0/m_b^0}{V_{cb}} \sim \delta \sim \frac{1}{5}$. There is
nothing in principle wrong with this, except that as $\theta$ is the
angle between two supposedly unrelated vectors, $\vec{a}$ and
$\vec{b}$, it might have been expected to be closer to unity.
Second, the smallness of $(m_c^0/m_t^0)/
(m_s^0/m_b^0)$ is not so much {\it explained} as {\it fit}, and,
in particular, a special choice of the form of $(I + \Delta_{d^c})
^{-\frac
{1}{2}}$ has to be made. ({\it Cf}. Eq. (24).) This choice
amounts to the statement that the vector $c_i$ (or $c_i'$)
points nearly in the `3' direction.
There is then a kind of preferred direction in family space,
in spite of our having eschewed family symmetry. (Though, on
the other hand, the `small numbers' involved in this fortuitous
alignment are only
of order $\frac{1}{5}$, whereas the intergenerational hierarchies
being explained involve ratios of $10^{-2}$ to $10^{-5}$.)
A third cost is that the exactness of the relation $m_b^0
\cong m_{\tau}^0$ is lost.
For example, the simple form given in Eq. (24) would lead
(for small $\epsilon$ and $c_{23}$) to $m_b^0 \cong \delta
(aTv')$ and $m_{\tau}^0 \cong \sqrt{\delta^2 \cos^2 \theta
+ \sin^2 \theta} (aTv')$. Since both $\delta$ and $\theta$
are of order $\frac{1}{5}$, $m_b^0/m_{\tau}^0$ can deviate
from unity by as much as $40 \%$. As we shall see in Section 3,
the form of $\Delta_{d^c}$ can be chosen to fit $m_b^0/m_{\tau}^0$.
But now it can only be claimed as a ``prediction" that the masses
of $b$ and $\tau$ are equal in a rough sense. Finally, because the
directions of $c_i$ and $c_i'$ are not arbitrary but must be
chosen to give $\Delta_{d^c}$ the required form, it turns out that
there is a hierarchy among the $c_{ij}$. In particular,
$c_{23}$ is somewhat larger than $c_{12}$ and $c_{13}$, and
therefore, while still small compared to the terms in which it
enters, it is not so small as to be negligible. In fact, the
Georgi-Jarlskog factors of 3 are significantly affected.
(See Section 3.)

\vspace{0.5cm}

\noindent
{\large\bf (d) The long version of the model}

While all of the costs of fitting $(m_c^0/m_t^0)/
(m_s^0/m_b^0)$ that had to be paid in the short version
of the model are relatively minor, they do somewhat take away from
the cleanness of the model and its explanatory power.
There is, however, another very beautiful way to break the proportionality
relation. This involves adding the terms $W_{{\rm adoint}}$ given
in Eq. (3) to the superpotential. To some extent this reduces the
economy
of the model, though as emphasized earlier the overall structure of
$W_{{\rm adjoint}}$ is parallel to that of $W_{{\rm spinor}}$ and
$W_{{\rm vector}}$. Moreover, by adding these terms one is enabled
not merely to fit but to {\it explain} the smallness of $(m_c^0/m_t^0)/
(m_s^0/m_b^0)$ in a rather elegant way, as will now be explained.
Finally, this explanation is achieved in such a way that the other
positive features of the model are virtually unaffected. None of
the costs that have to be paid in the short version have to be paid here.
This then is a much cleaner version of the model.

If one assumes that $\langle {\bf 5} ( \overline{{\bf 16}}_H) \rangle =
0$ (this will be seen to be desirable on other grounds in Section 5)
there is no additive contribution to the mass matrices (analogous
to the $c_{ij}$) coming from $W_{{\rm adjoint}}$. $W_{{\rm adjoint}}$
does introduce, however, new
multiplicative corrections of the form $(I + \Delta_f)^{-\frac{1}{2}}$,
where $f = u^c$, $u$, $d$, and $l^+$,
that reflect the mixing of the ${\bf 10} ({\bf 16}_i)$ with the
${\bf 10} ({\bf 45})$ and ${\bf 10} ({\bf 45}')$. This mixing comes
from the following terms contained in $W_{{\rm adjoint}}$:
\begin{equation}
\begin{array}{cc}
& \overline{{\bf 10}} ({\bf 45}) \left[ (f \langle {\bf 45}_H \rangle)
\cdot {\bf 10} ({\bf 45}') + \sum_i (e_i \langle {\bf 1}
(\overline{{\bf 16}}_H) \rangle ) \cdot {\bf 10} ({\bf 16}_i) \right] \\
& \\ + &
\overline{{\bf 10}} ({\bf 45}') \left[ (f \langle {\bf 45}_H \rangle)
\cdot {\bf 10} ({\bf 45}) + \sum_i (e_i' \langle {\bf 1}
(\overline{{\bf 16}}_H) \rangle ) \cdot {\bf 10} ({\bf 16}_i) \right].
\end{array}
\end{equation}

\noindent
The effect of this is to introduce factors of the form $(I +
\Delta_f)^{-\frac{1}{2}}$ into the mass matrices $U$, $D$, and
$L$ wherever they multiply an $SU(5)$ ${\bf 10}$ of fermions.
For example, the matrix $U_{ij}$ now takes the form

\begin{equation}
U = aTv (I + \Delta_{u^c})^{-\frac{1}{2}} \left( \begin{array}{ccc}
0 & 0 & 0 \\
0 & 0 & \frac{1}{5} \epsilon \sin \theta/N_u \\
0 & -(1+ \frac{4}{5} \epsilon) \sin \theta /N_{u^c} & -(1+ \frac{3}{5}
\epsilon)
\cos \theta / N_{u^c} N_u
\end{array} \right) (I + \Delta_u)^{-\frac{1}{2}}.
\end{equation}

\noindent
The matrices $\Delta_{u^c}$, $\Delta_u$, $\Delta_d$, and $\Delta_{l^+}$
all have the form
\begin{equation}
(\Delta_{f})_{ij} = E_i^* E_j + E_i^{* \prime} E_j^{\prime},
\end{equation}

\noindent
where
\begin{equation}
(E_1,E_2,E_3) \equiv \left| \frac{ \langle {\bf 1}
(\overline{{\bf 16}}_H) \rangle}{f \Omega Q_{f({\bf 45})}}
\right| (e_1,e_2,e_3/N_f),
\end{equation}

\noindent
and similarly for the definition of $E_i'$. (The expressions
for $\Delta_{d^c}$, $\Delta_{l^-}$, and $\Delta_{\nu}$ have already been
given in Eqs. (22) and (23).)

A point of crucial importance is that the expression for $\Delta_f$,
$f = u^c$, $u$, $d$, and $l^+$, given in Eqs. (27) and (28) has a factor of
$Q_{f({\bf 45})}$ in the denominator. This simply comes from the
VEV of the ${\bf 45}_H$ that appears in Eq. (25).
Note that this adjoint
is the same one that appears in $W_{{\rm spinor}}$ (rather than the
$\tilde{{\bf 45}}_H$ that appears in
$W_{{\rm vector}}$) and has its VEV in the direction $Q$. But it can
be seen from Eq. (3) that the generator $Q$ in $W_{{\rm adjoint}}$ is not
acting on fermions that are in a ${\bf 16}$, as in $W_{{\rm spinor}}$,
but on fermions that are in a ${\bf 45}$. What is the difference?
Using the relation $I_{3R} = -\frac{1}{10} X + \frac{3}{5} \left(
\frac{Y}{2} \right)$, one can rewrite Eq. (7) as
\begin{equation}
\begin{array}{c}
Q = \left( \frac{-1}{5} \right) X + \left( \frac{6}{5} (1+ \epsilon)
\right)
\frac{Y}{2}.
\end{array}
\end{equation}

\noindent
The $SU(5)$-singlet generator $X$ (as we have
normalized it) takes the value 1 on the representation
${\bf 10}({\bf 16})$, but
-4 on the representation ${\bf 10}({\bf 45})$. Thus
\begin{equation}
Q_{f({\bf 45})} = Q_{f({\bf 16})} + 1.
\end{equation}

\noindent
Thus the charges that enter into Eq. (28) are
\begin{equation}
\begin{array}{ccccl}
Q_{u({\bf 45})} & = & Q_{d({\bf 45})} & = & 1 + \frac{1}{5} \epsilon \\
& & Q_{u^c({\bf 45})} & = & - \frac{4}{5} \epsilon \\
& & Q_{l^+({\bf 45})} & = & 2 + \frac{6}{5} \epsilon.
\end{array}
\end{equation}

\noindent
Note the important fact that for $\left| \epsilon \right| \ll 1$
the {\it only} field in the ${\bf 45}$ that has a
small charge is the $u^c$. The remarkable consequence of this
is that the elements of $\Delta_{u^c}$ are of order $\frac{1}
{\epsilon^2}$ (see Eqs. (25), (26), and (29)), while all the other
$\Delta_f$ have elements that are not enhanced in this way.
In this long version of the model, then, the assumption can be
made that all the $\Delta_f$ are less than unity
so that the factors
$(I + \Delta_f)^{-\frac{1}{2}}$ are unimportant, {\it except}
for the matrix $\Delta_{u^c}$, which has large elements of
order $\frac{1}{\epsilon^2}$.

Because of the foregoing {\it only the matrix} $U$ {\it gets
substantially affected by the presence of} $W_{{\rm adjoint}}$.
Most of the successful relations of the model are then essentially
unaffected, since they follow from the forms of $D$ and $L$.
The quantities that depend strongly on the form of $U$ are $m_u$,
$m_c/m_t$, and the mixing angles. But, as the factor $(I +
\Delta_{u^c})^{-\frac{1}{2}}$ does not alter the fact that $U$
is rank 2, the smallness of $m_u$ is still explained. And because
the factor $(I +
\Delta_{u^c})^{-\frac{1}{2}}$ involves a transformation of the
{\it right-handed} quarks, the mixing angles are not strongly
affected. One might expect that the cancellation between the
contributions from the up and down sectors that makes $V_{cb}$
small would be seriously disrupted if $(I +
\Delta_{u^c})^{-\frac{1}{2}}$ were very different from the
identity matrix. It turns out, however, that $V_{cb}$ remains
of order $\epsilon$. In fact (see Appendix B for the derivation)
$V_{cb}^0 \cong \frac{2}{5} \epsilon \sin \theta \cos \theta
\left[ 1 + \frac{N}{2} \tan \theta \frac{(\vec{E} \times \vec{E}')_2}
{(\vec{E} \times \vec{E}')_3} \right]$.

{}From these considerations it is seen that the {\it only
prediction of the model that is substantially altered by
the addition of the term} $W_{{\rm adjoint}}$ {\it is the proportionality
relation for} $m_c^0/m_t^0$. At first glance one might expect that
the factor $(I +
\Delta_{u^c})^{-\frac{1}{2}}$ would be just as likely to
increase as to decrease the ratio $m_c^0/m_t^0$, for arbitrary
values of the parameters in $\Delta_{u^c}$. But, remarkably, this
proves not to be the case. In Appendix B we give an explicit proof which
shows that if the elements of $\Delta_{u^c}$ are large then
$m_c^0/m_t^0$ will be {\it suppressed} except for special directions
in parameter space. (Interestingly, this is a
consequence of the fact that $\Delta_{u^c}$ is a rank 2 matrix, as can be
seen from Eq. (27).  The ratio $(m_c^0/m_t^0)/(m_s^0/m_b^0)$ in this case is
given by
$$ {(m_c^0/m_t^0) \over (m_s^0/m_b^0)} \simeq
{ |\vec{E} \times \vec{E'}| \over |(\vec{E} \times \vec{E'})_3|^2 }
\sqrt{|E_1|^2+|E_1'|^2} \simeq {\cal O} (\epsilon) ,$$
from which is is clear that $(m_c^0/m_t^0)$ is
suppressed relative to $(m_s^0/m_b^0)$ for $|E_i| \gg 1$ (i.e.,
$\epsilon \ll 1$, Cf. Eq. (28)).  See Appendix B
for a derivation.
If it were an arbitrary matrix of rank one or rank three, then
$m_c^0/m_t^0$ would with equal likelihood be suppressed or
enhanced. This also is shown in Appendix B. These statements have
also been checked by numerical tests.)
Of course, if the elements of $\Delta_{u^c}$ are
small, $m_c^0/m_t^0$ will be only slightly affected one way or the other.
In particular, since the elements of $\Delta_{u^c}$ are of order
$\frac{1}{\epsilon^2}$ the ratio $m_c^0/m_t^0$ is suppressed by
a factor of order $\epsilon$ as shown above (see Appendix B),
which is what is needed to agree with experiment.

The surprising conclusion is that the addition of the terms
$W_{{\rm adjoint}}$ to the superpotential for generic values
of the parameters suppresses the ratio $m_c^0/m_t^0$ by order
$\epsilon$, while having only a minor effect otherwise.
The violation of the proportionality relation is therefore
not merely fit in some arbitrary way, but explained in a
group-theoretical way. In this version of the model,
the relation $\left| \epsilon \right| \ll 1$ plays a role in
explaining no less than five facts! Such economy of explanation
is not something that can be contrived.

\section{Numerical Fits}

{\large \bf (a) Long version}

In this Section we shall present the details of our numerical fits.
Let us first focus on the long version of the model.  In this version,
we can consistently set the multiplicative matrices $\Delta_u,
\Delta_d, \Delta_{d^c}, \Delta_{l^-}, \Delta_{\nu}$ and
$ \Delta_{l^+}$ (Cf. Eqs. (19)-(21))
all to zero.  The elements of $\Delta_{u^c}$ are enhanced by a factor
$1/|\epsilon|^2$ and therefore $\Delta_{u^c}$ cannot be ignored
(Cf. Eq. (26)).  In this limit we have the following approximate
analytic expressions for the mass ratios and mixing angles:
\begin{eqnarray}
m_t^0 & \simeq & {aTv \over N} {|(\vec{E} \times \vec{E'})_3| \over
|\vec{E} \times \vec{E'}|} \nonumber \\
{m_c^0 \over m_t^0} & \simeq & {N \over 5} \epsilon {\rm sin}^2\theta
{|\vec{E} \times \vec{E'}|\sqrt{E_1^2+E_1'^2}} \over {|(\vec{E} \times
\vec{E'})_3|^2} \nonumber \\
m_u^0/m_t^0 &\simeq & 0
\end{eqnarray}
\begin{eqnarray}
m_b^0 &\simeq & {aTv \over N} \nonumber \\
{m_s^2 \over m_b^0} &\simeq & {N \over 5} \epsilon {\rm sin}^2\theta
\nonumber \\
{m_d^0 m_s^0 \over m_b^{0^2}} & \simeq & c_{12}N^2(c_{12}{\rm
cos}\theta-c_{13}{\rm sin}\theta)
\end{eqnarray}
\begin{equation}
m_\tau^0 \simeq m_b^0,~ m_\mu^0 \simeq 3 m_s^0,~ m_e^0 \simeq {1 \over
3} m_d^0
\end{equation}
\begin{eqnarray}
V^0_{ub} &\simeq & c_{13} \nonumber \\
V^0_{cb} &\simeq& {2 \over 5} \epsilon {\rm cos}\theta{\rm sin}\theta
\left[1+{N \over 2} {(\vec{E}\times \vec{E'})_2 \over (\vec{E}\times
\vec{E'})_3}\right] \nonumber \\
V_{us}^0 &\simeq & \sqrt{m_d^0/m_s^0}
\left[{\rm cos}\theta-{c_{13}\over
c_{12}} {\rm sin}\theta\right]^{-1/2}
\end{eqnarray}

Note that in this version, the angle $\theta$ is of order one, as a
result, there is a significant correction to the expression
tan$\theta_c \simeq \sqrt{m_d^0/m_s^0}$.  However, the ratio
$c_{13}/c_{12}$ is nearly equal to $0.5$ in order to fit $V_{ub}$, and
with this value the correction factor in Eq. (25) is small for
a wide range of the angle $\theta$ as long as  $c_{13}$ and
$c_{12}$ have a relative negative sign. (For example, if
$c_{13}/c_{12}=-1/2$, the correction factor in Eq. (25) is 1.04
corresponding to $\theta = 60^0$.)
This fact is borne out in our numerical fits.

A good fit follows by choosing
\begin{eqnarray}
\epsilon &=& 0.15 i,~ \vec{e} = (1,1,0),~ \vec{e'} = (1, -1, 0)
\nonumber \\
T &=& 1,~ {\rm sin}\theta = {\rm cos}\theta ,~ {4 f \Omega \over 5
 \left \langle 45_H \right \rangle}  = 1.2 \nonumber \\
c_{12} &=& 0.0035,~ c_{13} = -0.002,~ c_{23} = 0.002
\end{eqnarray}

The resulting mass eigenvalues are
\begin{eqnarray}
(m_u^0, m_c^0, m_t^0) &=& [0, 3.64 \times 10^{-3}, 0.708] aTv \nonumber
\\
(m_d^0, m_s^0, m_b^0) &=& [8.54 \times 10^{-4}, 1.59 \times 10^{-2},
0.709] aTv' \nonumber \\
(m_e^0, m_\mu^0, m_\tau^0) &=& [2.98 \times 10^{-4}, 4.53 \times
10^{-2}, 0.707] aTv
\end{eqnarray}
The quark mixing angles have the values
\begin{equation}
V^0_{us} = 0.213, ~V_{ub}^0 = 0.002,~ V_{cb}^0 = 0.030
\end{equation}
Since the input parameters are all complex in general, there is
room for sufficient KM type CP violation.  In the specific
example given above, this CP violation is somewhat small, but for other
choices of input phases large enough CP violation can be obtained.

\noindent{\large \bf (b) Short version}

Here the fit is somewhat nontrivial since the same set of vectors
$c_i$ and $c_i'$ of Eq. (2) should generate
the antisymmetric contribution $c_{ij}$ of
Eq. (13) as well as the multiplicative factor $\Delta_{d^c} =
\Delta_{l^-}$ given in Eq. (19)-(20) needed to correct the
proportionality relation.  As already noted, this can be achieved
by choosing an approximate form of $\Delta_{d^c}$ as given in Eq.
(24).  However, $m_b^0$ and $m_\tau^0$ are now only approximately
equal for generic values of the model parameters,
to within 40\% or so.  We shall choose the parameters so that
$m_b^0 \simeq m_\tau^0$ remains good to within about 10\%.  The
angle $\theta$ in this case is of order $1/5$, so that the
correction to the relation tan$\theta_c \cong \sqrt{m_d^0/m_s^0}$ is
small.  We will also need $T \ll 1$ so that $N \simeq 1$.

The following choice of parameters gives a good fit:
\begin{eqnarray}
\vec{c} &=& ( 1, 0.7, 7), ~\vec{c'} = (0.75, 1.5, 4.5) \times 10^{-3}
\nonumber \\
c_{ij} &=& (c_i c_j'-c_i'c_j),~ {\rm sin}\theta = 0.22,~
\epsilon = 0.45i,~ T = 0.25
\end{eqnarray}
The masses and mixing angles with this choice are
\begin{eqnarray}
(m_u^0, m_s^0, m_t^0) &=& [0, 4.5 \times 10^{-3}, 1.0]aTv \nonumber\\
(m_d^0, m_s^0, m_b^0) &=& [1.7 \times 10^{-4}, 4.2 \times 10^{-3},
0.22]aTv' \nonumber \\
(m_e^0, m_\mu^0, m_\tau^0) &=& [5.2 \times 10^{-5}, 1.2 \times 10^{-2},
0.25]aTv'
\end{eqnarray}
\begin{equation}
V^0_{us} = 0.213,~ V^0_{ub} = 0.0025,~ V_{cb}^0 = 0.0295
\end{equation}
Note that these numbers correspond to the
ratios $m_\mu^0/m_s^0 = 2.82, m_d^0/m_e^0 = 3.26, m_c^0/m_t^0 = 1/225,
m_s^0/m_b^0 = 1/52, (m_c^0/m_t^0)/(m_s^0/m_b^0) = 1/4.3$,  all of which
are in
good agreement with data.

\section{Predictions}

\noindent
{\large\bf (a) $\tan \beta$}

If one neglects all the factors of $(I+ \Delta_f)^{-\frac{1}{2}}$,
then $v/v' \cong m_t^0/m_b^0 \cong m_c^0/m_s^0$. Of course, it
is the failure of the latter equality that requires one to assume that not
all of the mixing matrices are trivial. In the short version of the
model, the mass of the $b$ quark is suppressed by such effects
while the masses of $s$, $c$, and $t$ are little affected. Thus
$v/v' \cong m_c^0/m_s^0$ still holds but $m_t^0/m_b^0$ is larger.
If one assumes that $\langle \overline{{\bf 5}} ( {\bf 16}_H) \rangle
\ll \langle \overline{{\bf 5}} ({\bf 10}_H) \rangle$ (see the
discussion in Section 5(a) for why this is sensible) then $\tan \beta
\cong v/v'$ and there follows the unusual prediction that
$\tan \beta^0 \cong m_c^0/m_s^0$.  After renormalization group
corrections, tan$\beta$ at weak scale is $\approx 18$ (corresponding
to $m_t = 175 ~GeV$.)

In the long version of the model, the masses of $b$ and $s$ are
little affected by the mixing matrices, but both $m_c^0$ and
$m_t^0$ are suppressed. (See Appendix B.) However, $m_t^0$
is only suppressed by a factor of order unity, while $m_c^0$
is suppressed by a factor of order $\epsilon$. Thus (again
assuming $\tan \beta \cong v/v'$) one has the prediction
$\tan \beta \simeq m_t^0/m_b^0$.  (This relation is unaffected
by renormalization group running.)

\noindent
{\large\bf (b) Neutrino mixing angles}

The predictions for neutrinos will be discussed first in the long
version of the model where they are simpler. In the long version it
is assumed that all the $\Delta_f$ are somewhat less than unity
so that the factors $(I +
\Delta_f)^{-\frac{1}{2}}$ are close to unity and negligible,
except for $\Delta_{u^c}$, which is enhanced by the group-theoretical
effect discussed above. But obviously the factor $(I +
\Delta_{u^c})^{-\frac{1}{2}}$ does not affect the neutrino mixing angles
at all.

At first glance it might be thought that there could be no
predictions for neutrinos in either version of this model,
since the neutrino mass matrix, $M_{\nu}$, depends not only on the
Dirac neutrino mass matrix, $N$, which is known (after fitting
the charged fermion masses and mixing angles), but also on the
superheavy, Majorana mass matrix, $M_R$, which is unrelated
directly to the others and not known or predicted in this model.
\begin{equation}
M_{\nu} = N^T M_R^{-1} N.
\end{equation}

\noindent
However, ignoring those effects which produce $m_u \neq 0$,
the matrix $N$, like $U$, has rank 2, with vanishing first row and
column in the basis in which we have worked. (See Eq. (11).)
It is then clear that (in this limit) for {\it any} form
of $M_R^{-1}$, the matrix $M_{\nu}$ also has vanishing first
row and column. Thus, in diagonalizing $M_{\nu}$ no rotation
in the 1-2 or 1-3 planes is necessary, and the mixing angles
$V_{e {\mu}}$ and $V_{e {\tau}}$ come entirely from the diagonalization
of $L$, which is a known matrix, just as $V_{us}$ and $V_{ub}$
come from the diagonalization of $D$. Since the form of $L$ is similar to that
of $D$, there are relations between the lepton and quark mixing
angles. In particular, since $L_{33} \cong D_{33}$ and $L_{31}
= - D_{31}$, one has $V^{{\rm lepton}}_{13} \cong - (V_{{\rm KM}})_{13}$,
or
\begin{equation}
V_{e {\tau}} \cong - V_{ub}.
\end{equation}

\noindent
And for the same reason that $V_{us} \cong \sqrt{m_d^0/m_s^0}$,
\begin{equation}
V_{e {\mu}} \cong \sqrt{m_e^0/m_{\mu}^0}.
\end{equation}

It is likely that whatever effect makes $U$ a rank 3 matrix also
makes $N$ rank 3. Just as these effects are likely to correct
$\tan \theta_c$ by O$(m_u^0/m_c^0)$, one would expect that
the corrections to $V_{e {\mu}}$ and
$V_{e {\tau}}$ would be roughly of order $m_u^0/m_c^0 \approx 0.005$
and $m_u^0/m_t^0 \approx 10^{-5}$ respectively. But it is possible
also that the effects that produce $m_u$ act differently on the
neutrinos.

The mixing angle $V_{{\mu} {\tau}}$ cannot be predicted in the absence
of any information about $M_R$. However, in certain limits interesting
predictions arise. Simply multiplying out Eq. (42) one finds that

\begin{equation}
\begin{array}{ccl}
(M_{\nu})_{33} & = & (M_R^{-1})_{33} (1+ \frac{3}{5} \epsilon )^2
\cos^2 \theta/N^2 \\
& + & 2 (M_R^{-1})_{23} \frac{3}{5} \epsilon (1 + \frac
{3}{5}\epsilon ) \sin \theta \cos \theta/N \\
& + & (M_R^{-1})_{22} (\frac{3}{5}
\epsilon)^2 \sin^2 \theta, \\
& & \\
(M_{\nu})_{23} & = & (M_{\nu})_{32} = (M_R^{-1})_{33} (1 + \frac{3}{5}
\epsilon ) \sin \theta \cos \theta/ N^2 \\
& + & (M_R^{-1})_{23}
\frac{3}{5} \epsilon \sin^2 \theta/N, \\
& & \\
(M_{\nu})_{22} & = & (M_R^{-1})_{33} \sin^2 \theta/N^2. \\
\end{array}
\end{equation}

\noindent
And, of course, $(M_{\nu})_{1i} = (M_{\nu})_{i1} = 0$. From these
equations it follows that

\begin{equation}
\begin{array}{cccl}
& (M_R^{-1})_{33} & > & (M_R^{-1})_{23}, \;\; \epsilon(M_R^{-1})_{22} \\
& & & \\
\Rightarrow & V_{{\mu}{\tau}}^0 & = & -3 V_{cb}^0 \left[
1 + {\rm O} \left( \frac{(M_R^{-1})_{23}}{(M_R^{-1})_{33}} \tan \theta
\right) + {\rm O} (\epsilon) \right].
\end{array}
\end{equation}

\noindent
Another interesting case is
\begin{equation}
\begin{array}{cccl}
& (M_R^{-1})_{23} & > & (\epsilon \tan \theta)^{-1} (M_R^{-1})_{33},
\;\; \epsilon \tan \theta (M_R^{-1})_{22} \\
& & & \\
\Rightarrow & V_{{\mu}{\tau}}^0 & \cong & \frac{1}{2} \theta, \\
& m(\nu_{\mu})/m(\nu_{\tau}) & \cong & \frac{1}{4}.
\end{array}
\end{equation}

In the short version of the model the neutrino predictions
are complicated by the presence of the factors $(I +
\Delta_{l^-})^{-\frac{1}{2}} = (I +
\Delta_{\nu})^{-\frac{1}{2}}$ in Eqs. (20) and (21). These factors affect
the mixing of the left-handed charged leptons. In the absence of
these factors one would have the same predictions for neutrinos as in
the long version.

The matrix $\Delta_{l^-}$ is given by the expressions in
Eqs. (22) and (23) with $N_{d^c}$ replaced by $N_{l^-}$.
A satisfactory suppression of $m_b^0$ (and thus of
$(m_c^0/m_t^0)/(m_s^0/m_b^0)$) would be achieved if $c_1$, $c_2$,
$c_1'$, and $c_2'$ were very small compared to one, while
$c_3$ and/or $c_3'$ were larger than one. For simplicity of
discussion let us henceforth ignore $c_i'$ and just assume that
$c_3 > 1$. Then $(\Delta_{l^-})_{33} \cong \left| c_3 \right|^2
> 1$ and all other elements of $(\Delta_{l^-})$ are small, which
gives $(I + \Delta_{l^-})^{-\frac{1}{2}} \cong$ diag$(1,1,
1/\sqrt{1+ \left| c_3 \right|^2})$. This has the effect of
multiplying $L_{33}$ by $\frac{1}{\sqrt{1 + \left| c_3 \right|^2}}$,
and thus the rotation in the $(\mu_L^-, \tau_L^-)$-plane required
to diagonalize $L$ is not approximately $\tan \theta$ but
$\sqrt{1 + \left| c_3 \right|^2} \tan \theta \cong \frac{m_s^0/m_b^0}
{m_c^0/m_t^0} \tan \theta$, which numerically (see Section 3) is
of order unity. The short version, then, typically predicts
that $V_{{\mu}{\tau}}$ is {\it large}.

A second effect of the matrix $(I +
\Delta_{l^-})^{-\frac{1}{2}}$ arises from the fact that $c_1$ and
$c_2$ are not in general exactly zero. In fact, as was seen in Section 3,
one can fit the relation $m_b^0 \cong m_{\tau}^0$ by choosing
Re$(c_2/m_3) \cong -\frac{1}{2} \tan \theta$. Now, the presence of a
non-vanishing $c_1$, which would be natural to expect, leads to
non-vanishing (12) and (13) elements of $(I +
\Delta_{l^-})^{-\frac{1}{2}}$. These, in turn, contribute to
$V_{e {\mu}}$ and $V_{e {\tau}}$. In fact, $V_{e {\tau}}$ can be quite
large. Because of the presence of the unknown parameter $c_1$, there
are not independent predictions for $V_{e {\mu}}$ and $V_{e {\tau}}$,
but one prediction for $V_{e {\mu}}$ in terms of $V_{e {\tau}}$. For
$V_{e {\tau}}$ small, the prediction for $V_{e {\mu}}$ just goes
over to that for the long version of the model. More generally
\begin{equation}
V_{e {\mu}} \cong \sqrt{m_e^0/m_{\mu}^0} + {\rm O}(V_{e {\tau}}^0).
\end{equation}

\section{Technical Issues}

\noindent
{\large\bf (a) The Higgs sector: doublet-triplet splitting}

In $SO(10)$ SUSY GUTS the simplest way to naturally achieve the
doublet-triplet splitting of the Higgs fields that is required
for the gauge hierarchy is by the Dimopoulos-Wilczek mechanism.$^{12}$
The essential idea involves an adjoint of Higgs ($A_1$) whose
VEV is in the $B-L$ direction and a pair of fundamental Higgs
multiplets ($T_1$ and $T_2$). Consider the following Higgs
superpotential.
\begin{equation}
W_{2/3} = \lambda T_1 A_1 T_2 + M_2 (T_2)^2.
\end{equation}

\noindent
Since the $T_i$ are in ${\bf 10}$ representations, each contains
a ${\bf 5} + \overline{{\bf 5}}$ of $SU(5)$ that we will denote
${\bf 5}(T_i)$ and $\overline{{\bf 5}}(T_i)$. Then one has the
following mass matrix for those fields
\begin{equation}
(\overline{{\bf 5}}(T_1), \overline{{\bf 5}}(T_2))
\left( \begin{array}{cc} 0 & \lambda \langle A_1 \rangle \\
- \lambda \langle A_1 \rangle & M_2 \end{array} \right)
\left( \begin{array}{c} {\bf 5}(T_1) \\ {\bf 5}(T_2)
\end{array} \right).
\end{equation}

\noindent
If $\langle A_1 \rangle = a \cdot (B-L) + b \cdot (I_{3R})$,
then, in an obvious notation, one has for the masses of the
$SU(2)_L$-doublets and color-triplets contained in $T_i$
\begin{equation}
\begin{array}{ccl}
W_{{\rm mass}} & = & ( \overline{{\bf 2}}(T_1), \overline{{\bf 2}}
(T_2)) \left( \begin{array}{cc} 0 & \lambda b \\ - \lambda b & M_2
\end{array} \right) \left( \begin{array}{c} {\bf 2}(T_1) \\
{\bf 2}(T_2) \end{array} \right) \\
& & \\
& + & ( \overline{{\bf 3}}(T_1), \overline{{\bf 3}}
(T_2)) \left( \begin{array}{cc} 0 & \lambda a \\ - \lambda a & M_2
\end{array} \right) \left( \begin{array}{c} {\bf 3}(T_1) \\
{\bf 3}(T_2) \end{array} \right).
\end{array}
\end{equation}

\noindent
If $a \sim M_{{\rm GUT}} \sim M_2$, while $b=0$, there is
a massless pair of doublets, ${\bf 2}(T_1) + \overline{{\bf 2}}
(T_1)$, that play the role of $H + H'$ in the supersymmetric
standard model (SSM), while the other pair of doublets and all of
the triplets become superheavy. Moreover, if one assumes that
$T_2$ does not couple to light quarks and leptons the proton-decay
amplitude coming from the exchange of the color-triplet Higgsinos,
${\bf 3}(T_1) + \overline{{\bf 3}}(T_1)$, is proportional
to $M_2/(\lambda a)^2$. For the theory to be perturbative $\lambda$
cannot be large compared to unity, and so $\lambda a
\stackrel{_<}{_\sim} M_{{\rm GUT}}$. However, $M_2$ can be
somewhat smaller than $M_{{\rm GUT}}$, and this provides a means
by which the proton lifetime can be made consistent with
experiment.$^8$ (In the minimal $SU(5)$ SUSY GUT it is well-known
that the proton lifetime is only marginally consistent, if at all,
with experiment.$^9$)

In general, $b$ will not stay exactly zero if non-renormalizable
terms (induced by gravity, for example) are taken into account,
and, in fact, it is not a completely trivial matter to ensure
that nonrenormalizable terms do not destabilize it. How small must
$b$ remain to preserve the gauge hierarchy? As noted, the
proton-decay amplitude is proportional to $M_2/(\lambda a)^2$, which
must therefore be $\stackrel{_<}{_\sim} (10 M_{{\rm GUT}})^{-1}$.
The mass of the light Higgs doublets, on the other hand, is given by
$(\lambda b)^2/M_2$, and therefore can be written $m_H
\stackrel{_>}{_\sim} \left( \frac{b}{a} \right)^2 (10 \; M_{{\rm GUT}})$.
thus, it must be that $b/a \stackrel{_<}{_\sim} 10^{-7} \stackrel{_<}
{_\sim} (M_{{\rm GUT}}/M_{{\rm Pl}})^2$. That this can naturally be achieved
is shown in Ref. 16.

This simple picture of doublet-triplet splitting must be modified
in the context of the present model, since in addition to the
${\bf 10}_H$ appearing in $W_{{\rm spinor}}$, which is to be
identified with $T_1$, there is the ${\bf 16}_H$ appearing in
$W_{{\rm vector}}$, which also must break $SU(2)_L \times U(1)_Y$
and must as a consequence partially contain a light $\overline{{\bf 2}}$.
To be more exact, there is just a single light $\overline{{\bf 2}}$ that
gets a Weak-scale VEV and is a
linear combination of doublets in ${\bf 10}_H$ and ${\bf 16}_H$.
This raises two issues: (1) Can the Dimopoulos-Wilczek mechanism
for doublet-triplet splitting still work?$^{17}$, and (2) does the
mechanism for suppressing Higgsino-mediated
proton-decay still operate?

In the long version of the model there is also a $\overline{{\bf 16}}_H$
that is required {\it not} to get an $SU(2)_L \times U(1)_Y$-breaking
VEV. So issue (3) is whether this is natural. Finally, it is desirable
that $\langle \overline{{\bf 5}}({\bf 16}_H) \rangle/\langle
\overline{{\bf 5}}({\bf 10}_H) \rangle$ (which we shall define to
be $\tan \gamma$) be small for two reasons. First, it
would explain (see Eq. (13))
why the $c_{ij}$ are small, and, second, it would mean that
the light Higgs, $H'$,
of the SSM would be almost purely in the ${\bf 10}_H$, so that the ratio
$\langle {\bf 5}({\bf 10}_H) \rangle/ \langle \overline{{\bf 5}}
({\bf 10}_H) \rangle$, which is predicted in the model,
is just the empirically measurable parameter $\tan \beta$. Issue (4)
is whether the smallness of $\tan \gamma$ can be simply and naturally
achieved.

Let us denote ${\bf 10}_H$, ${\bf 16}_H$, and $\overline{{\bf 16}}_H$
by $T_1$, $C$, and $\overline{C}$. A satisfactory generalization$^{17}$
of Eq. (49) is
\begin{equation}
W_{2/3} = \lambda T_1 A_1 T_2 + M_2(T_2)^2 + \rho T_1 \overline{C}
\overline{C} + M_C \overline{C} C.
\end{equation}

\noindent
which gives
\begin{equation}
W_{{\rm mass}} = ( \overline{{\bf 5}}(T_1), \overline{{\bf 5}}(T_2),
\overline{{\bf 5}}(C)) \left( \begin{array}{ccc}
0 & \lambda \langle A_1 \rangle & \rho \langle \overline{C} \rangle \\
- \lambda \langle A_1 \rangle & M_2 & 0 \\ 0 & 0 & M_C
\end{array} \right) \left( \begin{array}{c}
{\bf 5}(T_1) \\ {\bf 5}(T_2) \\ {\bf 5}(\overline{C}) \end{array}
\right).
\end{equation}

\noindent
If $b=0$ (that is, if $\langle A_1 \rangle = a(B-L)$), then the
matrix for the doublets has one massless eigenvalue. The massless
${\bf 2}$ is purely in $T_1$, while the massless $\overline{{\bf 2}}$
is a linear combination of $\overline{{\bf 2}}(T_1)$ and $\overline
{{\bf 2}}(C)$:
\begin{equation}
\overline{{\bf 2}}_{{\rm light}} = \cos \gamma \overline{{\bf 2}}(T_1)
+ \sin \gamma \overline{{\bf 2}}(C),
\end{equation}

\noindent
with
\begin{equation}
\tan \gamma = - \rho \langle \overline{C} \rangle/ M_C.
\end{equation}

\noindent
Since the orthogonal, superheavy doublet, $\overline{{\bf 2}}_{{\rm heavy}}
= - \sin \gamma \overline{{\bf 2}}(T_1) + \cos \gamma \overline{{\bf 2}}
(C)$, must have vanishing VEV, it follows that
$\langle \overline{{\bf 2}}(C) \rangle / \langle \overline{{\bf 2}}
(T_1) \rangle \equiv \langle \overline{{\bf 5}}({\bf 16}_H) \rangle/
\langle \overline{\bf 5}({\bf 10}_H) \rangle$ $ = \tan \gamma$.
Thus by making the ratio $\rho \langle \overline{C} \rangle/ M_C$ small
one ensures that issue (4) raised above is satisfactorily resolved.
One possibility, discussed later, is that the term $\rho T_1
\overline{C} \overline{C}$ arises from a higher-dimension operator,
so that $\rho \sim {\rm O}(M_{{\rm GUT}}/M_{{\rm Pl}})$.

It is necessary, as before, that $b/a \stackrel{_<}{_\sim} 10^{-7}$
to preserve the gauge hierarchy (issue (1)), but it is no longer
sufficient. It is also necessary that the lower left entry
in Eq. (53 ),
call it $x$, which connects $\overline{{\bf 5}}(C)$ to ${\bf 5}(T_1)$
be extremely small. To be precise, it must be that $x \rho \langle
\overline{C} \rangle/M_C \stackrel{_<}{_\sim} m_W$, or, in other
words, $x \stackrel{_<}{_\sim} \cot \gamma m_W$.
It should be noted that this would also automatically ensure that
$\langle {\bf 5}(\overline{{\bf 16}}_H) \rangle \cong 0$, which
resolves issue (3). For $\langle {\bf 5}(\overline{{\bf 16}}_H)
\rangle/ \langle {\bf 5} ({\bf 10}_H) \rangle \equiv
\langle {\bf 2}(\overline{C}) \rangle/
\langle {\bf 2}(T_1) \rangle = x/M_C \stackrel{_<}{_\sim}
m_W/\rho \langle \overline{C} \rangle \sim m_W/M_{{\rm GUT}}.$

The remaining question (issue 2) is whether Higgsino-mediated
proton decay can still be suppressed by making $M_2$ somewhat
smaller than $M_{{\rm GUT}}$. This is easily seen to be the case
from the form of Eq. (53). If $M_2$ is set to zero, then the colored
Higgsino ${\bf 3}(T_1)$ (which is by assumption the only ${\bf 3}$
to couple to the light quarks and leptons) only has a mass connecting
it to $\overline{{\bf 3}}(T_2)$, which does not mix with the
$\overline{{\bf 3}}(T_1)$ and $\overline{{\bf 3}}(C)$ that couple to
the light quarks and leptons. Thus the Higgsino-mediated proton-decay
amplitude is proportional to $M_2$ as in the simpler case of Eq. (49).
Note also that there is a group theoretical suppression of order
$\epsilon$ in the rate for proton decay.$^2$

In order for this simple scenario to work naturally, we have seen
that several conditions must be satisfied. (a) As before, $b/a$ must
be $\stackrel{_<}{_\sim} 10^{-7}$. (b) For $x$ to be sufficiently small
the coefficient of any effective $T_1 C C$ term must be less than
about $m_W/M_{{\rm GUT}}$. And (c) the $T_1 \overline{C} \overline{C}$
term should have a small (O($M_{{\rm GUT}}/M_{{\rm Pl}}$)) coefficient.
In the next subsection we will see that a realistic superpotential
satisfying these criteria can be constructed.

\vspace{0.5cm}

\noindent
{\large\bf (b) The Higgs sector: the breaking of $SO(10)$}

The model of quarks and leptons requires the existence of
adjoints with VEVs in the $I_{3R}$ and $X$ directions. We will denote
these by $A_2$ and $A_3$, respectively. The Dimopoulos-Wilczek mechanism
requires the existence of an adjoint whose VEV is in the $B-L$
direction, which we have been denoting $A_1$. (As above, we will denote
the $\overline{{\bf 16}}_H$ and ${\bf 16}_H$ by $\overline{C}$ and
$C$, and the two ${\bf 10}_H$ by $T_1$ and $T_2$.) A satisfactory
form for the superpotential is

\begin{equation}
\begin{array}{ccl}
W_{{\rm Higgs}} & = & {\rm tr}(A_1)^4/M_{{\rm Pl}} + M_{A1}(A_1)^2 \\
& + & {\rm tr}(A_2)^4/M_{{\rm Pl}} + M_{A2}(A_2)^2 \\
& + & M_C \overline{C} C + \frac{M_C}{M_{{\rm Pl}}^2} \overline{C}
(A_3)^2 C + M_{A3}(A_3)^2 \\
& + & {\rm tr}(A_1 A_2 A_3) \\
& + & T_1 A_1 T_2 + M_2 (T_2)^2 + \frac{R}{M_{{\rm Pl}}} T_1
\overline{C} \overline{C}.
\end{array}
\end{equation}

\noindent
We have not written the dimensionless coefficients, which are assumed
to be of order unity. This as a hybrid of the forms proposed in
Ref. 16.

The form $W(A) = {\rm tr}(A^4)/M_{{\rm Pl}} + M(A)^2$ has as possible
solutions $A \propto X$, $B-L$, and $I_{3R}$, with $\left| \langle
A \rangle \right| \sim \sqrt{M_{{\rm Pl}} M}$. Thus $M_{A1}$ and
$M_{A2}$ in Eq. (56) must be of order $M_{{\rm GUT}}^2/M_{{\rm Pl}}$.

The terms involving $A_3$, $\overline{C}$, and $C$ can be shown to have
a solution in which the VEVs of these fields are in $SU(5)$-singlet
directions. The magnitude of $\langle A_3 \rangle$ is determined by
the $F_C$ and $F_{\overline{C}}$ equations to be O($M_{{\rm Pl}}$).
This means that $SO(10)$ is broken to $SU(5) \times U(1)_X$ near
the Planck scale. The fractional mass splittings within $SU(5)$
multiplets will then be O($M_{{\rm GUT}}/M_{{\rm Pl}}$) and many
of the threshold corrections to $\sin^2 \theta_W$ will be suppressed
by that small ratio. (See Ref. 16.) On the other hand, it is
desirable that the VEV of $\overline{C}$ be of order $M_{{\rm GUT}}$
because of the role it plays in $W_{{\rm adjoint}}$. The scale of
$\langle \overline{C} \rangle$ and $\langle C \rangle$ is determined
by the $F_{A3}$ equation to be O($\sqrt{M_{A3}/M_C}M_{{\rm Pl}}$).

The term tr$(A_1 A_2 A_3)$, by linking the $A_1$, $A_2$, and $A_3$
sectors, ensures that there are no goldstone modes, while at the same time
not destabilizing the VEVs of the $A_i$. (See the first paper of
Ref. 16 for a discussion of this term.)

The terms involving $T_i$ have already been discussed in the last
subsection. If $\langle R \rangle \sim M_{{\rm GUT}}$, then $\tan \gamma
\sim M_{{\rm GUT}}/M_{{\rm Pl}}$ as desired.

\vspace{0.5cm}

\noindent
{\large\bf (c) Discrete symmetries}

The essential core of the model of quark and lepton masses consists
of the Yukawa terms given in Eqs. (1) -- (3). In particular, the
``root model" is defined by the terms in Eq. (1). The most
important feature of that set of terms is that there is no direct
$g_{ij} {\bf 16}_i {\bf 16}_j {\bf 10}_H$ coupling. This structure
can be explained by a very simple $Z_2$ parity under which the
Higgs fields, ${\bf 10}_H \equiv T_1$ and ${\bf 45}_H \equiv A_2$,
and the families, ${\bf 16}_i \equiv F_i$, are odd, while the
extra real representations of matter fields, ${\bf 16} \equiv F$ and
$\overline{{\bf 16}} \equiv \overline{F}$, are even.

The structure of the complete set of Yukawa terms given in Eqs. (1) -- (3)
can be ensured by a $Z_3 \times Z_2$ symmetry, where $Z_2$ is a matter
parity under which ``matter fields" are odd and Higgs fields are even,
and $Z_3$ is a symmetry which acts on the fields as shown in Table I.
(The other extra real representations of matter are denoted
${\bf 10} \equiv T$, ${\bf 10}' \equiv T'$, ${\bf 45} \equiv A$,
and ${\bf 45}' \equiv A'$.)

This particular symmetry allows in addition to the terms in Eqs. (1)
-- (3) a few extra harmless terms (namely $\overline{F} \overline{F} T_1$,
$\overline{F} T \overline{C}$, and $\overline{F} T' \overline{C}$).

A realistic Higgs superpotential (that is, one which completely
breaks $SO(10)$, avoids goldstone modes, preserves the unification
of couplings, naturally achieves doublet-triplet splitting and sufficient
suppression of Higgsino-mediated proton decay, and gives adjoint VEVs
in the desired directions) can be constructed. And symmetries can be found
that ensure its structure and the stability of the gauge hierarchy
against possible Planck-scale effects. This has been shown in Ref. 16.

It remains to show that symmetries can be found which render
natural the full model, including both the Higgs and Yukawa
parts of the superpotential. This is done in Appendix C. It should be
emphasized that most of the technical difficulty of making the model
natural has to do with the Higgs sector, and that the problem is
not made significantly more difficult by the particular Yukawa
structure assumed.

\section{Variants of the Root Model}

\noindent
{\large\bf (a) A variant of the root model with $\epsilon \cong -\frac{5}{4}$}

As explained in Section 2(a), the root model consisting of the terms
in $W_{{\rm spinor}}$ predicts that
\begin{equation}
\frac{m_{\mu}^0}{m_s^0} \cong \frac{Q_{l^+} Q_{l^-}}{Q_{d^c}q_d}
= 3 \left| \frac{1+ \frac{6}{5} \epsilon}{1+ \frac{2}{5} \epsilon}
\right|.
\end{equation}

\noindent
There are {\it two} values of $\epsilon$ which give the Georgi-Jarlskog
result $m_{\mu}^0/m_s^0 \cong 3$, namely $\epsilon \cong 0$ and
$\epsilon \cong - \frac{5}{4}$. The first corresponds to $Q \cong
I_{3R}$ and gives the model proposed in Ref. 1 and studied in detail
in previous sections. The second value gives an interesting variant
that we shall now discuss briefly.

With $\epsilon = -\frac{5}{4} + \delta$, $\left| \delta \right| \ll 1$,
the mass matrices take the forms (see Eqs. (8) --(10)
\begin{equation}
U_0 \cong aTv \left( \begin{array}{ccc}
0 & 0 & 0 \\
0 & 0 & (-\frac{1}{4} + \frac{1}{5} \delta) \sin \theta \\
0 & \frac{4}{5} \delta \sin \theta  & (-\frac{1}{4} -\frac{3}{5} \delta)
\cos \theta
\end{array} \right),
\end{equation}

\begin{equation}
D_0 \cong aTv' \left( \begin{array}{ccc}
0 & 0 & 0 \\
0 & 0 & (\frac{1}{4} + \frac{1}{5} \delta) \sin \theta \\
0 & (\frac{1}{2} + \frac{2}{5} \delta) \sin \theta  &
(\frac{1}{4} + \frac{3}{5} \delta)
\cos \theta
\end{array} \right),
\end{equation}

\begin{equation}
L_0 \cong aTv' \left( \begin{array}{ccc}
0 & 0 & 0 \\
0 & 0 & (\frac{3}{4} - \frac{3}{5} \delta) \sin \theta \\
0 & (-\frac{1}{2} + \frac{6}{5} \delta) \sin \theta &
(\frac{1}{4} + \frac{3}{5} \delta)
\cos \theta
\end{array} \right).
\end{equation}

There is one immediately apparent explanatory success: $(m_c^0/m_t^0)/
(m_s^0/m_b^0)$ $ = {\rm O}(\delta)$. However, the contributions to
$V_{cb}$ from the up and down sectors no longer nearly cancel and
$V_{cb}^0 \approx \sqrt{m_s^0/m_b^0}$. The situation is thus the reverse
of the model with $\left| \epsilon \right| \ll 1$. There the prediction
for $V_{cb}^0$ is good and that for $(m_c^0/m_t^0)/(m_s^0/m_b^0)$ is
bad and has to be corrected by another mechanism. Here $(m_c^0/m_t^0)
/(m_s^0/m_b^0)$ works well, but $V_{cb}^0$ has to be corrected by some other
mechanism.
The question arises: How could $V_{cb}^0$ be corrected without disturbing
the Georgi-Jarlskog relation? One possibility is the following.

All the mass matrices coming from $W_{{\rm spinor}}$ have the form
\begin{equation}
M_0 = m_0 \left( \begin{array}{ccc}
0 & 0 & 0 \\
0 & 0 & Q \sin \theta \\
0 & Q^c \sin \theta & (Q^c + Q) \cos \theta
\end{array}
\right).
\end{equation}

\noindent
This means that
\begin{equation}
\frac{m_{\mu}^0}{m_s^0} \cong \frac{m_{\mu}^0 m_{\tau}^0}
{m_s^0 m_b^0} = \frac{\det_{23} L_0}{\det_{23} D_0}
= \frac{Q_{l^+} Q_{l^-}}{Q_{d^c} Q_d},
\end{equation}

\noindent
while
\begin{equation}
V_{cb}^0 \cong \left( \frac{Q_{d^c}}{Q_{d^c} + Q_d}
- \frac{Q_{u^c}}{Q_{u^c} + Q_u} \right) \tan \theta.
\end{equation}

\noindent
Consider adding to each matrix a contribution
\begin{equation}
\Delta M = m_0 \left( \begin{array}{ccc}
0 & 0 & 0 \\
0 &  \Delta \cdot \sin^2 \theta & \Delta \cdot \sin \theta
\cos \theta \\
0 & \Delta \cdot \sin \theta \cos \theta & \Delta \cdot
\cos^2 \theta
\end{array}
\right),
\end{equation}

\noindent
with $\Delta$ the same at least for $D$ and $L$. Then
$D_{33}$ and $L_{33}$ remain equal, and so, for small
$\theta$, $m_b^0 \cong m_{\tau}^0$ remains valid and
$\frac{m_{\mu}^0}{m_s^0} \cong \frac{\det_{23} L}{\det_{23} D}$
still holds. But $\det_{23} (M_0 + \Delta M) = (\Delta \sin^2 \theta)
((Q^c + Q) \cos \theta + \Delta \cos^2 \theta) - (Q \sin \theta
+ \Delta \sin \theta \cos \theta)(Q^c \sin \theta + \Delta
\sin \theta \cos \theta) = - Q Q^c \sin^2 \theta$, which is
unaffected by the addition of $\Delta M$! Thus, the Georgi-Jarlskog
relation remains good. On the other hand, for small $\theta$,
\begin{equation}
V_{cb}^0 \cong \left[ \frac{Q_{d^c} + \Delta \cos \theta}
{(Q_{d^c} + Q_d) + \Delta \cos \theta} - \frac {Q_{u^c} +
\Delta \cos \theta}{(Q_{u^c} + Q_u) + \Delta \cos \theta}
\right] \tan \theta,
\end{equation}

\noindent
which is very greatly affected by the addition of $\Delta M$, and
can easily be made small.

How can such a $\Delta M$ arise? In the original root model,
defined by $W_{{\rm spinor}}$, the $SU(2)_L \times U(1)_Y$-breaking
term connects $\sum_i \hat{a}_i {\bf 16}_i = \sin \theta {\bf 16}_2
+ \cos \theta {\bf 16}_3$ with ${\bf 16}$, which in turn mixes
with $\sum_i \hat{b}_i {\bf 16}_i = {\bf 16}_3$. This gives the form
of $M_0$. Clearly, if there were effectively a term of the form
$\sum_{ij} \hat{a}_i \hat{a}_j {\bf 16}_i {\bf 16}_j$ it would
give a contribution of the form $\Delta M$. Such a term can arise
in a modified version of the root model. Imagine that instead
of a simple vectorlike pair of family, there were
two such pairs: $\overline{{\bf 16}} + {\bf 16} + \overline{{\bf 16}}'
+ {\bf 16}'$. Consider the superpotential $(M \overline{{\bf 16}}
{\bf 16} + \sum_i b_i \overline{{\bf 16}} {\bf 16}_i {\bf 45}_H)$
$+ (M' \overline{{\bf 16}}' {\bf 16}' + \sum_i a_i \overline{{\bf 16}}'
{\bf 16}_i {\bf 1}_H)$ $+ g {\bf 16} \; {\bf 16}' {\bf 10}_H + h {\bf 16}'
{\bf 16}' {\bf 10}_H$. Clearly the ${\bf 16}$ and ${\bf 16}'$ mix with
$\sum_i \hat{b}_i {\bf 16}_i$ and $\sum_i \hat{a}_i {\bf 16}_i$,
respectively.  Then the last two terms give, effectively, contributions
of the form $M_0$ and $\Delta M$.

This variant root model has several features which make it seem less
attractive than the model with $\epsilon \cong 0$. In the latter model
the choice of small $\epsilon$ explains three facts (the smallness of $V_{cb}$,
the 2nd to 3rd generation hierarchy, and $m_{\mu}^0/m_s^0 \cong 3$)
and also helps explain the suppression of Higgsino-mediated proton decay.
(In the long version it also plays a crucial role in explaining the
suppression of $m_c^0/m_t^0$.) In the variant root model the choice
of $\epsilon$ explains only two facts (the smallness of $m_c^0/m_t^0$
and $m_{\mu}^0/m_s^0 \cong 3$). Secondly, the variant model is less
economical by virtue of the introduction of the extra pair of family
and anti-family, and involves a reintroduction of a family symmetry
of sorts, since the $\overline{{\bf 16}}' + {\bf 16}'$ must be
distinguished from the $\overline{{\bf 16}} + {\bf 16}$. Finally,
the point $\epsilon = -\frac{5}{4}$, unlike the point $\epsilon = 0$,
does not seem to correspond to a
group-theoretically interesting direction. Nevertheless, perhaps some
of the ideas in this section are capable of further interesting development.

\vspace{0.5cm}

\noindent
{\large\bf (b) A variant way to suppress $m_c^0/m_t^0$}

In Sections 2(c) and (d) two ways to break the proportionality
relation $m_c^0/m_t^0 = m_s^0/m_b^0$ based on a multiplicative correction
to the mass matrices were described. They give the two versions of
the model. Another possibility is a mechanism based on an additive
correction to the mass matrices given in Eqs. (8) -- (11). (We are
now assuming again that $\epsilon \cong 0$.) Adding something to $D$
to increase $m_s^0/m_b^0$ without disturbing $V_{cb}^0$ or the
Georgi-Jarlskog relation seems difficult if not impossible.
On the other hand, a small additive correction to $U$ could
approximately cancel off the small $U_{23}$ element without
significantly affecting $U_{32}$ or $U_{33}$. Then $m_c^0/m_t^0$
would be suppressed without affecting $V_{cb}^0$ or any of the relations
that come from the forms of $D$ and $L$.

There is only one difficulty with this idea: such an additive contribution
to $U$ would in general make $U$ be rank 3 and, possibly, make $m_u$ too
large. Since $\Delta U_{23} \cong - U_{0,23} \cong (m_s^0 m_t^0/m_b^0)
/\sin \theta \gg m_u^0$, the form of $\Delta U$ must be such that
$\Delta U_{11} \ll \Delta U_{23}$.

One way to ensure that $\Delta U$ has a form that maintains the rank 2
nature of $U_0$ is by extending the root model ($W_{{\rm spinor}}$)
in a way analogous to what was described in Section 6(a).
Introduce, as there, two pairs of family plus anti-family instead
of one. Consider the superpotential $W_{{\rm spinor}}' =
(M \overline{{\bf 16}} {\bf 16} + \sum_i b_i \overline{{\bf 16}}
{\bf 16}_i {\bf 45}_H)$ $ + (M' \overline{{\bf 16}}' {\bf 16}' +
\sum_i a_i \overline{{\bf 16}}' {\bf 16}_i {\bf 1}_H)$
$+ g {\bf 16} \; {\bf 16}' {\bf 10}_H + \sum_i f_i {\bf 16}_i
{\bf 16}' \overline{{\bf 16}}_H \overline{{\bf 16}}_H$. The term
$g {\bf 16} \; {\bf 16}' {\bf 10}_H$ gives effectively a contribution
of the same form as Eq. (4). The last term contributes to $U$ (if
$\langle {\bf 1}(\overline{{\bf 16}}_H) \rangle$ and
$\langle {\bf 5}(\overline{{\bf 16}}_H) \rangle$ are both non-zero),
but not to $D$ or $L$. The sum of the last two terms effectively
involves only two linear combinations of ${\bf 16}$'s, and therefore
the resulting total $U$ is still rank 2.

The first generation can still be given mass by adding $W_{{\rm vector}}$.
The present variant would dispense with need for $W_{{\rm adjoint}}$,
however. The main drawback compared to the long version of the model
described in Section 2(d) is that the predictions of neutrino mixing
angles are lost because of the extra parameters $f_i$.

\section*{Appendix A}

\noindent
{\large\bf Derivation of exact mass matrices}

\noindent
Consider the matrix
\begin{equation}
{\cal M} = \left( \begin{array}{cc}
m_0 & M' \\ m & M \end{array} \right),
\end{equation}

\noindent
where $M$ and $M'$ contain elements of order $M_{{\rm GUT}}$
and $m$ and $m_0$ contain only Weak scale entries.
${\cal M}$ may be block-diagonalized as follows.
\begin{equation}
{\cal U}_R {\cal M} {\cal U}_L^{\dag} =
\left( \begin{array}{cc}
(I + x x^{\dag})^{-\frac{1}{2}}(m_0 -M'M^{-1}m) & 0 \\
0 & (M^{\dag} M + M^{\prime \dag} M')^{\frac{1}{2}}
\end{array} \right),
\end{equation}

\noindent
where
\begin{equation}
{\cal U}_R = \left( \begin{array}{cc}
(I + x x^{\dag})^{-\frac{1}{2}} & 0 \\
0 & (M^{\dag} M + M^{\prime \dag} M')^{-\frac{1}{2}}
\end{array} \right)
\left( \begin{array}{cc} I & -x \\
M^{\prime \dag} & M^{\dag}
\end{array} \right),
\end{equation}

\noindent
and
\begin{equation}
{\cal U}_L = \left( \begin{array}{cc}
I & -(m_0^{\dag} M' + m^{\dag} M)(M^{\dag} M + M^{\prime \dag} M')^{-1} \\
(M^{\dag} M + M^{\prime \dag} M')^{-1} (M^{\prime \dag} m_0
+ M^{\dag} m) & I \end{array} \right).
\end{equation}

\noindent
Here $x \equiv M' M^{-1}$. Terms of order $(M_{{\rm Weak}}/M_{{\rm GUT}})^2$
have been dropped. This see-saw form will now be applied to the
case of the mass matrix of the charge-$(-\frac{1}{3})$ quarks
in the short version of the model.

We will define the superheavy linear combination of ${\bf 16}$'s
to be ${\bf 16}' \equiv \frac{1}{\sqrt{1 + T^2 |Q|^2}} [ {\bf 16}
+ T Q {\bf 16}_3 ]$, and the light orthogonal combination
to be ${\bf 16}'_3 \equiv \frac{1}{\sqrt{1 + T^2 |Q|^2}} [
- T Q {\bf 16} + {\bf 16}_3 ]$. Then
\begin{equation}
\begin{array}{ccl}
{\bf 16} & = & \frac{1}{\sqrt{1 + T^2 |Q|^2}} [ {\bf 16}'
- T Q {\bf 16}'_3 ], \\
{\bf 16}_3 & = & \frac{1}{\sqrt{1 + T^2 |Q|^2}} [
+ T Q {\bf 16}' + {\bf 16}'_3 ].
\end{array}
\end{equation}

\noindent
Define $d_1$, $d_2$, $d'_3$, $d'$, $\overline{d}^c$, $g$, and $g'$
to be the charge-$(-\frac{1}{3})$ quarks in the ${\bf 16}_1$,
${\bf 16}_2$, ${\bf 16}'_3$, ${\bf 16}'$, $\overline{{\bf 16}}$,
${\bf 10}$, and ${\bf 10}'$, respectively; and define
$d_1^c$, $d_2^c$, $d_3^{\prime c}$, $d^{\prime c}$, $\overline{d}$,
$g^c$, and $g^{\prime c}$ to be the charge-$(\frac{1}{3})$
antiquarks in the same representations. Then by substituting
Eq. (70) into Eqs. (1) and (2), and
restricting attention to the down-type quarks, one gets the
following $7 \times 7$ mass matrix
\begin{equation}
W_{{\rm mass}} =
(d_1^c,d_2^c,d_3^{\prime c}, d^{\prime c}, \overline{d}, g^c,
g^{\prime c}) \cdot
\left( \begin{array}{cc}
m_0 & M' \\ m & M \end{array} \right)
\left( \begin{array}{c}
d_1 \\ d_2 \\ d_3' \\ d' \\ \overline{d}^c \\ g \\ g'
\end{array} \right),
\end{equation}

\noindent
where
\begin{equation}
m_0 =
\left( \begin{array}{ccc}
0 & 0 & 0 \\
0 & 0 & -aTv' Q_d s_{\theta}/N_d \\
0 & -aTv' Q_{d^c} s_{\theta}/N_{d^c} & -aTv'(Q_{d^c} + Q_d) c_{\theta}
/N_{d^c} N_d
\end{array} \right),
\end{equation}

\begin{equation}
m =
\left( \begin{array}{ccc}
0 & a v' s_{\theta}/N_{d^c} & a v' (1 - T^2 Q_{d^c} Q_d)c_{\theta}
/N_{d^c} N_d \\
0 & 0 & 0 \\
c_1 \tilde{v}' & c_2 \tilde{v}' & c_3 \tilde{v}'/N_{d^c} \\
c_1' \tilde{v}' & c_2' \tilde{v}' & c_3' \tilde{v}'/N_{d^c}
\end{array} \right),
\end{equation}

\begin{equation}
M' =
\left( \begin{array}{cccc}
0 & 0 & c_1 v_R & c_1' v_R \\
a v' s_{\theta}/N_d & 0 & c_2 v_R & c_2' v_R \\
a v' (1 - T^2 Q_{d^c} Q_d) c_{\theta}/N_{d^c} N_d & 0
& c_3 v_R/N_{d^c} & c_3' v_R/N_{d^c}
\end{array} \right),
\end{equation}

\begin{equation}
M =
\left( \begin{array}{cccc}
a T v' (Q_{d^c} + Q_d) c_{\theta}/N_{d^c} N_d &
M N_{d^c} & c_3 T Q_{d^c} v_R/N_{d^c} &
c_3' T Q_{d^c} v_R/N_{d^c} \\
M N_{d^c} & 0 & 0 & 0 \\
c_3 T Q_d \tilde{v}'/N_d & 0 & 0 & -d\langle \tilde{{\bf 45}}_H \rangle \\
c_3' T Q_d \tilde{v}'/N_d & 0  & +d\langle \tilde{{\bf 45}}_H \rangle &
0  \end{array} \right),
\end{equation}

\noindent
and where $v_R \equiv \langle {\bf 1}({\bf 16}_H) \rangle$,
$\tilde{v}' \equiv \langle \overline{{\bf 5}}({\bf 16}_H) \rangle$.
$v' \equiv \langle \overline{{\bf 5}}({\bf 10}_H) \rangle$,
$c_{\theta} \equiv \cos \theta$, and $s_{\theta} \equiv \sin \theta$.
Then by Eq. (57)
\begin{equation}
D_0 = m_0,
\end{equation}
\noindent
and
\begin{equation}
D = (I + x x^{\dag})^{-\frac{1}{2}} (m_0 - M' M^{-1} m).
\end{equation}

\noindent
Comparing with Eq. (19) one sees that
\begin{equation}
\Delta_{d^c} = x x^{\dag} = (M' M^{-1}) (M' M^{-1})^{\dag}.
\end{equation}

\noindent
Multiplying out this expression using Eqs. (74) and (75)
(one can neglect the terms of order $v'$ and $\tilde{v}'$
in $M'$ and $M$) gives the result in Eqs. (22) and (23).
And multiplying out
the expression $M' M^{-1} m$ gives the flavor-antisymmetric
piece in Eq. (13).

The same kind of calculation gives the mixing matrices
in the long version of the model as well.

\section*{Appendix B}

\noindent
{\large\bf Suppression of $m_c/m_t$ in long version of model}

\noindent
{}From Eq. (27) one has
\begin{equation}
\Delta_{u^c} = \vec{E}^* \vec{E}^T + \vec{E}^{\prime *} \vec{E}^{\prime T}.
\end{equation}

\noindent
Define
\begin{equation}
\vec{P} \equiv \vec{E} \times \vec{E}' \left[ 1 +
\frac{\left| \vec{E} \right|^2 + \left| \vec{E}' \right|^2}
{\left| \vec{E} \times \vec{E}' \right|^2} \right|
^{\frac{1}{2}}.
\end{equation}

\noindent
So that
\begin{equation}
P^2 \equiv \left| \vec{P} \right|^2 = \left| \vec{E} \right|^2
+ \left| \vec{E}' \right|^2 + \left| \vec{E} \times \vec{E}' \right|^2.
\end{equation}

\noindent
We will assume that $\left| \vec{E} \right|$ and $\left| \vec{E}'
\right|$ are large compared to one (in fact, of order $1/\epsilon$),
so that there
is a hierarchy: $P \sim E^2, E^{\prime 2} \gg E,E' \gg 1$.
Further, define
\begin{equation}
\begin{array}{ccl}
\vec{F} & \equiv & \hat{P}^* \times \vec{E}, \\
\vec{F}' & \equiv & \hat{P}^* \times \vec{E}'.
\end{array}
\end{equation}

\noindent
Then one can write an exact expression for the square of the
mixing matrix $(I + \Delta_{u^c})^{-\frac{1}{2}}$:
\begin{equation}
(I + \Delta_{u^c})^{-1} = \frac{1}{1 + \left| P \right|^2}
\left[ I + \vec{F} \vec{F}^{\dag} + \vec{F}' \vec{F}^{\prime \dag}
+ \vec{P} \vec{P}^{\dag} \right].
\end{equation}

\noindent
This expression is easily checked by going to a particular
basis. Since it is in a rotationally invariant form, it is
true in any basis.

One can write the full matrix $U$ as
\begin{equation}
U \cong (I + \Delta_{u^c})^{-\frac{1}{2}} U_0,
\end{equation}

\noindent
with
\begin{equation}
U_0 \cong \left( \begin{array}{ccc}
0 & 0 & 0 \\
0 & 0 & \eta \\
0 & \sin \theta & \cos \theta
\end{array} \right) = (\vec{0}, \vec{A}, \vec{B})
\end{equation}

\noindent
where
\begin{equation}
\eta \equiv \frac{1}{5} \epsilon \sin \theta,
\end{equation}

\begin{equation}
\vec{A} \equiv \left( \begin{array}{c} 0 \\ 0 \\
\sin \theta \end{array} \right),
\end{equation}

\begin{equation}
\vec{B} \equiv \left( \begin{array}{c} 0 \\ \eta \\
\cos \theta \end{array} \right).
\end{equation}

\noindent
Then the matrix $U_0$ has eigenvalues
\begin{equation}
\begin{array}{ccl}
m_{t,0} & \cong & 1, \\
m_{c,0} & \cong & - \eta \sin \theta = \left|
\vec{A} \times \vec{B} \right|.
\end{array}
\end{equation}

\noindent
To find the eigenvalues of $U$ one considers
\begin{equation}
\begin{array}{ccl}
U^{\dag} U & = & U_0^{\dag} (I + \Delta_{u^c})^{-1} U_0 \\
& = & \frac{1}{1+ \left| P \right|^2}
\left( \begin{array}{c} \vec{0} \\ \vec{A}^{\dag} \\
\vec{B}^{\dag} \end{array} \right) \cdot
( I + \vec{F} \vec{F}^{\dag} + \vec{F}' \vec{F}^{\prime \dag}
+ \vec{P} \vec{P}^{\dag} ) \cdot ( \vec{0}, \vec{A}, \vec{B})
\end{array}
\end{equation}

\noindent
Then
\begin{equation}
\begin{array}{ccl}
(U^{\dag} U)_{33} & = & \frac{1}{1+ \left| P \right|^2}
(\left| \vec{A}^{\dag} \cdot \vec{P} \right|^2 +
\left| \vec{A}^{\dag} \cdot \vec{F} \right|^2 +
\left| \vec{A}^{\dag} \cdot \vec{F}' \right|^2 +
\left| \vec{A} \right|^2) \\
(U^{\dag} U)_{22} & = & \frac{1}{1+ \left| P \right|^2}
(\left| \vec{B}^{\dag} \cdot \vec{P} \right|^2 +
\left| \vec{B}^{\dag} \cdot \vec{F} \right|^2 +
\left| \vec{B}^{\dag} \cdot \vec{F}' \right|^2 +
\left| \vec{B} \right|^2) \\
(U^{\dag} U)_{23} & = & \frac{1}{1+ \left| P \right|^2}
(\vec{A}^{\dag} \cdot \vec{P} \vec{P}^{\dag} \cdot \vec{B} +
\vec{A}^{\dag} \cdot \vec{F} \vec{F}^{\dag} \cdot \vec{B} +
\vec{A}^{\dag} \cdot \vec{F}' \vec{F}'^{\dag} \cdot \vec{B} +
\vec{A}^{\dag} \cdot \vec{B}).
\end{array}
\end{equation}

\noindent
The first row and column are zero in our approximation.
One then has
\begin{equation}
\begin{array}{ccl}
m_t^2/m_{t,0}^2 \cong m_t^2 & \cong & {\rm tr} (U^{\dag} U) \\
& \cong & \frac{1}{1+ \left| P \right|^2} (\left| \vec{A}^{\dag}
\cdot \vec{P} \right|^2 + \left| \vec{B}^{\dag}
\cdot \vec{P} \right|^2) \\
& \cong & \frac{\left| P_3 \right|^2}{1 + \left| P \right|^2},
\end{array}
\end{equation}

\noindent
where we are keeping only the leading terms.
And
\begin{equation}
\begin{array}{cc}
m_c^2 m_t^2 \cong \det_{23} U^{\dag} U. & \\
& \\
\end{array}
\end{equation}
\noindent
Keeping only the leading terms (recalling the hierarchy
$P \ll F,F' \ll 1$) one has after some algebra
\begin{equation}
m_c^2 m_t^2  \cong  \frac{1}{(1+ \left| P \right|^2)^2}
\left[ \left| \vec{A}^{\dag} \cdot \vec{P}
\vec{B}^{\dag} \cdot \vec{F} -
\vec{B}^{\dag} \cdot \vec{P}
\vec{A}^{\dag} \cdot \vec{F} \right|^2
+ (F \rightarrow F') \right].
\end{equation}

\noindent
The crucial point is that the leading terms in the brackets,
which are O$(P^4)$, cancel leaving O($P^2$).
Further manipulations give
\begin{equation}
\begin{array}{ccl}
m_c^2 m_t^2 & = & \frac{1}{(1+ \left| P \right|^2)^2}
\left[ \left| (\vec{A} \times \vec{B})^{\dag} \cdot
(\vec{P} \times \vec{F}) \right|^2 + (F \rightarrow F') \right] \\
& & \\
& = & \frac{\left| P \right|^2}{(1+ \left| P \right|^2)^2}
\left[ \left| (\vec{A} \times \vec{B})^{\dag} \cdot \vec{E}
\right|^2 + (E \rightarrow E') \right] \\
& & \\
& = & \left| \vec{A} \times \vec{B} \right|^2
\frac{\left| P \right|^2}{(1+ \left| P \right|^2)^2}
(\left| E_1 \right|^2 + \left| E'_1 \right|^2) \\
& & \\
& \cong & m_{c,0}^2 m_{t,0}^2
\frac{\left| P \right|^2}{(1+ \left| P \right|^2)^2}
(\left| E_1 \right|^2 + \left| E'_1 \right|^2).
\end{array}
\end{equation}

\noindent
Or
\begin{equation}
\frac{m_c m_t}{m_{c,0} m_{t,0}} \cong \frac{\left| P \right|}
{1 + \left| P \right|^2} \sqrt{\left| E_1 \right|^2 +
\left| E'_1 \right|^2 },
\end{equation}

\noindent
and, since
$\frac{m_t^2}{m_{t,0}^2} \cong \frac{ \left| P_3 \right|^2}
{1 + \left| P \right|^2}$, the suppression factor is given
by
\begin{equation}
\frac{m_c/m_t}{m_{c,0}/m_{t,0}} \cong
\frac{\left| P \right|}{\left| P_3 \right|^2} \sqrt{ \left|
E_1 \right|^2 + \left| E'_1 \right|^2}.
\end{equation}

\noindent
Using the definition of $\vec{P}$
\begin{equation}
\frac{m_c/m_t}{m_{c,0}/m_{t,0}} \cong
\frac{\left| \vec{E} \times \vec{E}' \right|}
{\left| (\vec{E} \times \vec{E}')_3 \right|^2} \sqrt{ \left|
E_1 \right|^2 + \left| E'_1 \right|^2}.
\end{equation}

\noindent
This is the desired result. Notice that for large $E$ and $E'$
this goes as $1/\min (E,E')$, unless $\vec{E} \times
\vec{E}'$ happens to point nearly along the 3 direction.
In other words, generically the ratio $m_c/m_t$ is
{\it suppressed}. Since $E$, and $E'$ are of order $\frac{1}{\epsilon}$,
$m_c/m_t$ is suppressed by a factor of O$(\epsilon)$.

\vspace{0.5cm}

\noindent
{\large\bf Cases of $\Delta =$ rank 1 or rank 3}

\noindent
The curious fact that $m_c/m_t$ is generally suppressed
is a result of the fact that $\Delta_{u^c}$ is rank 2.
We can show this by doing the analogous calculation
for the cases where $\Delta$ is rank 1 and rank 3.
If $\Delta$ is rank 1 it can be written
\begin{equation}
\Delta = \vec{E}^* \vec{E}^T.
\end{equation}

\noindent
Then
\begin{equation}
(I + \Delta)^{-1} = I - \frac{1}{1 + \left| E \right|^2}
\vec{E}^* \vec{E}^T.
\end{equation}

\noindent
Parallelling the calculation for the rank-2 case closely
one finds
\begin{equation}
\frac{m_c/m_t}{m_{c,0} m_{t,0}} \cong
\frac{\sqrt{1 + \left| E_1 \right|^2} \sqrt{1 + \left|
\vec{E} \right|^2}} {1 + \left| E_1 \right|^2
+ \left| E_2 \right|^2}.
\end{equation}

\noindent
This is not small {\it unless} $\vec{E}$ happens to
be nearly in the 2 direction, which agrees with
intuitive expectation.

If $\Delta$ is rank 3, then $(I + \Delta)^{-1}$ is just an arbitrary
3-by-3 matrix which we will call $M$. A calculation similar
to the preceding two gives
\begin{equation}
\frac{m_c/m_t}{m_{c,0} m_{t,0}} \cong
\frac{\sqrt{M_{33} M_{22} - M_{32}^2}}{M_{33}}.
\end{equation}

\noindent
Again, this is not small except for particular special
choices of $\Delta$.

\vspace{0.5cm}

\noindent
{\large\bf A calculation of $V_{cb}^0$}

Let the transformation of the left-handed charge-$(\frac{2}{3})$
quarks required to diagonalize $U$ be

\begin{equation}
V_U = \left( \begin{array}{ccc} 1 & 0 & 0 \\
0 & c & s^* \\ 0 & -s & c \end{array} \right),
\end{equation}

\noindent
and the matrix required to diagonalize $U_0$ be the same
with $s \rightarrow s_0$ and $c \rightarrow c_0$.
Similarly, let the transformation of the left-handed
charge-$\frac{-1}{3}$ quarks required to diagonalize $D$
be of the same form with $s \rightarrow s'$ and $c \rightarrow c'$.
We take $c$, $c'$, and $c_0$ to be real and $s$, $s'$, and $s_0$
to be complex. Then
\begin{equation}
V_{cb}^0 \cong sc'-s'c
\end{equation}

\noindent
The primed quantities are easy to find: $s'/c' \cong D_{32}/D_{33}
\cong (1 - \frac{1}{5}
\epsilon) \tan \theta$ as can be seen from Eq. (15). (We ignore
the effects of $c_{23}$ and the $N_f$.) This gives
\begin{equation}
s' \cong (\sin \theta(1 - \frac{1}{5} \epsilon_R \cos^2 \theta),
-\frac{1}{5} \epsilon_I \sin \theta).
\end{equation}

\noindent
And $s_0 \cong (U_0)_{32}/(U_0)_{33}$ which is given
by the same expressions with $\epsilon \rightarrow
- \epsilon$, as can be seen from Eq. (14).
To find $s$ we can use the expressions derived for $U^{\dag} U$
earlier in this Appendix:

\begin{equation}
\frac{2 \left| s \right| c}{c^2 - \left| s \right|^2}
= \frac{2 \left| (U^{\dag} U)_{32} \right|}
{(U^{\dag} U))_{33} - (U^{\dag} U)_{22}},
\end{equation}

\begin{equation}
\arg s = \arg (U^{\dag} U)_{32}.
\end{equation}

\noindent
This gives, after multiplying out the expressions for
the elements of $U^{\dag} U$ given above,
\begin{equation}
\frac{2 \left| s \right| c}{c^2 - \left| s \right|^2}
= \frac{2 \left| s_0 \right|(c_0 + {\rm Re}K)}{c_0^2
- \left| s_0 \right|^2 + 2 c_0 {\rm Re} K} + {\rm O}( \eta^2),
\end{equation}

\noindent
and
\begin{equation}
\arg s = \arg s_0 - {\rm Im} K/c_0,
\end{equation}

\noindent
where
\begin{equation}
K \equiv \eta \left[ \frac{P_3 P_2^* + F_3 F_2^* + F'_3 F^{\prime *}_2}
{\left| P_3 \right|^2 + \left| F_3 \right|^2 +
\left| F'_3 \right|^2 + 1} \right].
\end{equation}

\noindent
Using $\eta \equiv \frac{1}{5} \epsilon \sin \theta$ and
the assumption that $P \gg F,F' \gg 1$, one obtains
\begin{equation}
K \cong \frac{1}{5} \epsilon \sin \theta (P_2^*/P_3^*).
\end{equation}

\noindent
These expressions can be combined to give after straightforward
algebra
\begin{equation}
V_{cb}^0 \cong \sin \theta \cos \theta (\frac{2}{5} \epsilon
- K/\cos \theta),
\end{equation}

\noindent
or
\begin{equation}
V_{cb}^0 \cong \frac{2}{5} \epsilon \sin \theta \cos \theta
\left[ 1 - \frac{1}{2} \tan \theta \left( \frac{P_2}{P_3} \right)^*
\right].
\end{equation}

\vspace{0.5cm}

\section*{Appendix C}

In order to find a symmetry that makes the superpotential of
Eq. (56) natural and stable it is necessary to introduce some
singlet superfields. This can be shown as follows. With the terms
tr$(A_1)^4$ and $(A_1)^2$, $A_1$ cannot transform non-trivially
except under a $Z_2$. (Similarly for $A_2$ and $A_3$.) But it
is crucial for the doublet-triplet splitting that $(T_2)^2$ be
allowed, while $(T_1)^2$ is forbidden, which implies a non-trivial
relative transformation of these two fields and hence of $A_1$ because
of the presence of the term $T_1 A_1 T_2$. Thus the terms tr$(A_1)^4$
and $(A_1)^2$ must be replaced with some form that allows a
non-trivial transformation of $A_1$. The simplest possibility is
to insert singlet superfields with non-trivial transformation
properties. For example, $\phi_1 {\rm tr} (A_1)^4/M_{{\rm Pl}}^2$.
For similar reasons it is convenient to introduce singlets into other
terms as well.

There is no point in trying to find the most elegant or simplest combination
of symmetries that works. Rather here it will only be shown that {\it some}
symmetry can be found. The easiest way to do this is to restrict the search
to a single $U(1)$ symmetry, and to introduce singlet fields where convenient
to make a needed term allowed. No attempt has been made to economize
on these singlets.

Consider, then, the following set of fields. Its $U(1)$ charge
is given in parentheses after the name of the field.
Adjoint Higgs: $A_1( -a_2 -a_3)$, $A_2(a_2)$, $A_3(a_3)$;
Fundamental Higgs: $T_1(t_1)$, $T_2(t_2$; Spinor Higgs:
$\overline{C}(-c)$, $C(c-x)$; Singlet Higgs: $P(-t_1 -t_2 +a_2 +a_3)$,
$Q(- 2t_2/3)$, $R(r)$, $S(s)$, $X(x)$, $Y(y)$; Matter Spinors:
$F_i(e)$, $F(-e -t_1)$, $\overline{F}(e + t_1 - s)$;
Matter Fundamentals: $T(-a_3/2 -y/2)$, $T'(-a_3/2 -y/2)$;
Matter Adjoints: $A(-a_2/2)$, $A'(-a_2/2)$. These charges are not
all independent, but satisfy $c = (a_3 - a_2 + 2x + y)/4$,
and $t_1 = -a_3/2 -3a_2/2 + x - y/2 + s$. With these charge
assignments the following terms are allowed:
\begin{equation}
\begin{array}{ccl}
W & = & {\rm tr}(A_1)^4 \phi_1/M_{{\rm Pl}}^2 + (A_1)^2 \tilde{\phi}_1 \\
& + &  {\rm tr}(A_2)^4 \phi_2/M_{{\rm Pl}}^2 + (A_2)^2 \tilde{\phi}_2 \\
& + & X \overline{C} C + X \overline{C} C (A_3 \phi_3)^2/M_{{\rm Pl}}^4
+ (A_3 \phi_3)^2/M_3 \\
& + & {\rm tr} (A_1 A_2 A_3) \\
& + & T_1 A_1 T_2 (P/M_{{\rm Pl}}) + (T_2)^2(Q^3/ M_{{\rm Pl}}^2)
+ T_1 \overline{C} \overline{C}(R/M_{{\rm Pl}}) \\
& + & [ s(\overline{F} F) + (\overline{F} F_i) A_2 + (F F_i) T_1 \\
& & + Y( T T') A_3/ M_{{\rm Pl}} + (T F_i)C + (T' F_i)C \\
& & + (A A') A_2 + (A F_i) \overline{C} + (A' F_i) \overline{C} ].
\end{array}
\end{equation}

\noindent
All the terms have coefficients of order unity that are not shown.
The VEVs of the fields are of the following orders of magnitude.
$\langle A_1 \rangle, \langle A_2 \rangle, \langle
\overline{C} \rangle, \langle C \rangle \sim M_{{\rm GUT}}$,
$\langle A_3 \rangle \sim M_{{\rm Pl}}$, $\langle X \rangle,
\langle Y \rangle, \langle P \rangle,\langle Q \rangle,
\langle R \rangle, \langle S \rangle \sim M_{{\rm GUT}}$,
$\langle \phi_i \rangle \sim M_{{\rm Pl}}$, and $\langle
\tilde{\phi}_i \rangle \sim M_{{\rm GUT}}/M_{{\rm Pl}}$.
(Thus $1/M_3$ must be of order $M_{{\rm GUT}}^3/M_{{\rm Pl}}^4$.)

The most dangerous terms for the hierarchy are those that lead effectively
to $T_1CC$ or to linear terms for $A_1$. The lowest term involving
$T_1CC$ is $T_1CC[A_2 \phi_3 R \overline{Y} ]/M_{{\rm Pl}}^4$
(assuming a field $\overline{Y}$ with opposite quantum numbers to
$Y$ exists). This gives effectively $(M_{{\rm GUT}}/M_{{\rm Pl}})^3
T_1CC$, which is sufficiently suppressed. The lowest dimension terms
linear in $A_1$ (that would destabilize $\langle A_1 \rangle$; note
that tr$(A_1 A_2 A_3)$ does not) are $A_1 (X \overline{C} C)
(R S \overline{Y})/M_{{\rm Pl}}^4$ and $(A_1 A_2 A_3)(X \overline{C}
C)/M_{{\rm Pl}}^3$. The first is harmless, but the second gives
effectively $(M_{{\rm GUT}}^4/M_{{\rm Pl}}^2)A_1$. Since the
mass of $A_1$ is necessarily of order $M_{{\rm GUT}}^2/M_{{\rm Pl}}$,
one has $\delta \langle A_1 \rangle / \langle A_1 \rangle \sim
M_{{\rm GUT}}/M_{{\rm Pl}}$. What is needed is $10^{-7}$, so that
this dangerous term must be assumed to have a dimensionless coefficient
that is of order $10^{-4}$. All other potentially dangerous terms are
sufficiently suppressed if one assumes that all terms are suppressed
only by dimensionally appropriate powers of the Planck mass.

\section*{References}
\begin{enumerate}
\item K.S. Babu and S.M. Barr, Phys. Rev. Lett. {\bf 75}, 2088 (1995).
\item K.S. Babu and S.M. Barr, BA-95-21 (hep-ph/9506261).
\item S.M. Barr, Phys. Rev. {\bf D24}, 1895 (1981);
Phys. Rev. Lett. {\bf 64}, 353 (1990).
\item Z. Berezhiani, Phys. Lett. {\bf 129B}, 99 (1983); B.S.
Balakrishna,
A. Kagan and R.N. Mohapatra, Phys. Lett. {\bf 205B}, 345 (1988);
J.C. Pati, Phys. Lett. {\bf B228}, 228 (1989);
K.S. Babu and E. Ma, Mod. Phys. Lett. {\bf A4}, 1975 (1989).
\item A. Buras, J. Ellis, M.K. Gaillard, and D.V. Nanopoulos,
Nucl. Phys. {\bf B135}, 66 (1978).
\item H. Georgi and C. Jarlskog, Phys. Lett. {\bf 86B}, 297 (1979).
\item H. Fritzsch, Nucl. Phys. {\bf B155}, 189 (1979).
\item G.D. Coughlan, G.G. Ross, R. Holman, P. Ramond, M. Ruiz-Altaba,
and J.W.F. Valle, Phys. Lett. {\bf 158B}, 401 (1985); J. Hisano,
H. Murayama, and T. Yanagida, Nucl. Phys. {\bf B402}, 46 (1993);
K.S. Babu and S.M. Barr, Phys. Rev. {\bf D48}, 5354 (1993).
\item N. Sakai and T. Yanagida, Nucl. Phys. {\bf B197} 533 (1982);
S. Weinberg, Phys. Rev. {\bf D26}, 287 (1982); S. Dimopoulos, S. Raby
andF. Wilczek, Phys. Lett.{\bf 112B}, 133 (1982);
J. Ellis, D.V.
Nanopoulos, and S. Rudaz, Nucl. Phys. {\bf B202}, 43 (1982);
P. Nath, A.H. Chamseddine, and R. Arnowitt, Phys. Rev. {\bf D32},
2348 (1985); R. Arnowitt and P. Nath, Phys. Rev. {\bf D49}, 1479
(1994).  For a review see J. Hisano, H. Murayama, and
T. Yanagida, Nucl. Phys. {\bf B402}, 46 (1993).
\item M. Srednicki, Nucl. Phys. {\bf B202}, 327 (1982).
\item K.S. Babu and S.M. Barr, Phys. Rev. {\bf D51}, 2463 (1995).
\item S. Dimopoulos and F. Wilczek, NSF-ITP-82-07, August 1981 and
Proceedings of Erice Summer School, ed. by A. Zichichi (1981);
K.S. Babu and S.M. Barr, Phys. Rev. {\bf D48},
5354 (1993); Phys. Rev. {\bf D50}, 3529 (1994).
\item S. Weinberg, in {\it Festschrift for I.I. Rabi}, Trans. N.Y.
Acad. Sci. Ser. II {\bf 38}, 185 (1977); F. Wilczek and A. Zee,
Phys. Lett. {\bf 70B}, 418 (1977); H. Fritzsch, Phys. Lett.
{\bf 70B}, 436 (1977).
\item R. Gatto, G. Sartori, and M. Tonin, Phys. Lett. {\bf 28B},
128 (1968); R.J. Oakes, Phys. Lett. {\bf 29B}, 683 (1969).
\item L. Wolfenstein, Phys. Rev. Lett. {\bf 41}, 1945 (1984).
\item K.S. Babu and S.M. Barr, Ref. 12 and Ref. 11.
\item K.S. Babu and R.N. Mohapatra, Phys. Rev. Lett.
{\bf 74}, 2418 (1995).
\end{enumerate}

\newpage

\noindent
{\large\bf Table I:}

$$
\begin{array}{c|c|c|c|c|c|c|c||c|c|c|c|c|c}
{\rm Matter:} & F_i & F & \overline{F} & T & T' & A & A'
& {\rm Higgs:} & A_2 & A_3 & T_1 & C & \overline{C} \\
\hline
Z_3 & 1 & z^2 & z & z^2 & z^2 & z & z & Z_3 & z & z^2 &
z^2 & z & z^2
\end{array}
$$

\noindent
{\large\bf Figure Captions}

\vspace{1cm}

\begin{description}
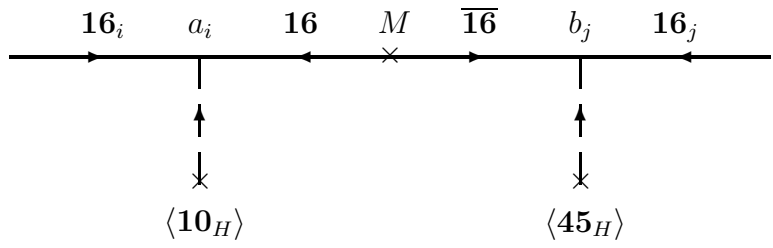
\item[Fig.1:] The $W_{spinor}$ contribution to light fermion masses.
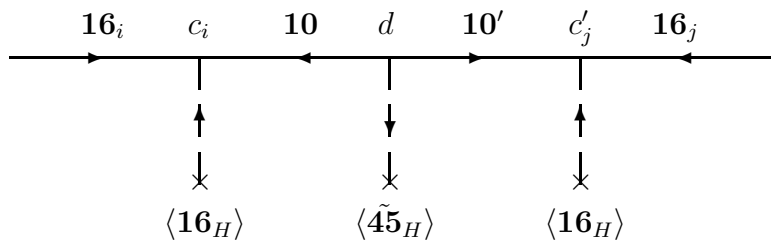
\item[Fig.2:] The $W_{vector}$ contribution to light fermion masses.
\end{description}

\newpage

\begin{picture}(360,216)
\thicklines
\put(36,144){\vector(1,0){36}}
\put(72,144){\line(1,0){72}}
\put(180,144){\vector(-1,0){36}}
\put(180,144){\vector(1,0){36}}
\put(216,144){\line(1,0){72}}
\put(324,144){\vector(-1,0){36}}
\put(108,96){\line(0,1){12}}
\put(108,114){\vector(0,1){12}}
\put(108,132){\line(0,1){12}}
\put(252,96){\line(0,1){12}}
\put(252,114){\vector(0,1){12}}
\put(252,132){\line(0,1){12}}
\put(175.5,142){$\times$}
\put(103.5,94){$\times$}
\put(247.5,94){$\times$}
\put(94.5,78){$\langle {\bf 10}_H \rangle$}
\put(238.5,78){$\langle {\bf 45}_H \rangle$}
\put(63,153){${\bf 16}_i$}
\put(103.5,153){$a_i$}
\put(139.5,153){${\bf 16}$}
\put(175.5,153){$M$}
\put(207,153){$\overline{{\bf 16}}$}
\put(247.5,153){$b_j$}
\put(279,153){${\bf 16}_j$}
\put(162,42){{\bf Fig. 1}}
\end{picture}

\begin{picture}(360,216)
\thicklines
\put(36,144){\vector(1,0){36}}
\put(72,144){\line(1,0){72}}
\put(180,144){\vector(-1,0){36}}
\put(180,144){\vector(1,0){36}}
\put(216,144){\line(1,0){72}}
\put(324,144){\vector(-1,0){36}}
\put(108,96){\line(0,1){12}}
\put(108,114){\vector(0,1){12}}
\put(108,132){\line(0,1){12}}
\put(252,96){\line(0,1){12}}
\put(252,114){\vector(0,1){12}}
\put(252,132){\line(0,1){12}}
\put(180,96){\line(0,1){12}}
\put(180,126){\vector(0,-1){12}}
\put(180,132){\line(0,1){12}}
\put(175.5,94){$\times$}
\put(103.5,94){$\times$}
\put(247.5,94){$\times$}
\put(94.5,78){$\langle {\bf 16}_H \rangle$}
\put(238.5,78){$\langle {\bf 16}_H \rangle$}
\put(166.5,78){$\langle \tilde{{\bf 45}}_H \rangle$}
\put(63,153){${\bf 16}_i$}
\put(103.5,153){$c_i$}
\put(139.5,153){${\bf 10}$}
\put(175.5,153){$d$}
\put(207,153){${\bf 10}'$}
\put(247.5,153){$c_j'$}
\put(279,153){${\bf 16}_j$}
\put(162,42){{\bf Fig. 2}}
\end{picture}

\end{document}